\documentclass[fleqn,usenatbib]{mnras}

\usepackage[T1]{fontenc}

\DeclareRobustCommand{\VAN}[3]{#2}
\let\VANthebibliography\thebibliography
\def\thebibliography{\DeclareRobustCommand{\VAN}[3]{##3}\VANthebibliography}

\usepackage{epsfig}
\usepackage{graphicx}	
\usepackage{amsmath}	
\usepackage{amssymb}	
\usepackage[T2A]{fontenc}
\usepackage{ wasysym }
\usepackage{multirow}
\usepackage{booktabs}
\usepackage{tabularx}
\usepackage{ulem}
\usepackage[dvipsnames]{xcolor}
\usepackage{tablefootnote}

\newcommand{\SII}{[S\,{\sevensize II}]\ }
\newcommand{\OII}{[O\,{\sevensize II}]\ }
\newcommand{\OIII}{[O\,{\sevensize III}]\ }
\newcommand{\ArIII}{[Ar\,{\sevensize III}]\ }
\newcommand{\NeIII}{[Ne\,{\sevensize III}]\ }

\newcommand{\NII}{[N\,{\sevensize II}]\ }
\newcommand{\HII}{H\,{\sevensize II}\ }
\newcommand{\HeII}{He\,{\sevensize II}\ }
\newcommand{\HeI}{He\,{\sevensize I}\ }

\newcommand{\NIIHa}{[N\,{\sevensize II}]/H$\alpha$}
\newcommand{\OIIIHb}{[O\,{\sevensize III}]/H$\beta$}
\newcommand{\Ha}{H$\alpha$\ }
\newcommand{\Hb}{H$\beta$\ }
\newcommand{\kms}{\,\mbox{km}\,\mbox{s}^{-1}}

\newcommand{\HST}{\textit{HST}\ }
\newcommand{\Msun}{\ensuremath{\rm M_\odot}}
\newcommand{\Lsun}{\ensuremath{\rm L_\odot}}

\newcommand{\be}{\begin{equation}}
\newcommand{\ee}{\end{equation}}

\newcommand{\Cloudy}{\textsc{Cloudy}} 

\usepackage{xcolor}

\renewcommand\fbox{\fcolorbox{red}{white}}

\title[The nitrogen-rich  massive star in NGC\,4068]{Unveiling  the nitrogen-rich massive star  in the metal-poor galaxy NGC\,4068}

\author[A. D. Yarovova et al.]{
Anastasiya D. Yarovova,$^{1}$\thanks{E-mail: yaan.ph@gmail.com}
Oleg V. Egorov,$^{2,1}$\thanks{E-mail: oleg.egorov@uni-heidelberg.de}, 
Alexei V. Moiseev$^{3,1}$ \\
\\
{\LARGE \rm and Olga V. Maryeva$^{4,3}$}\\
\\
$^{1}$ Lomonosov Moscow State University, Sternberg Astronomical Institute,
Universitetsky pr. 13, Moscow 119234, Russia
\\
$^{2}$ Astronomisches Rechen-Institut, Zentrum f\"{u}r Astronomie der Universit\"{a}t Heidelberg, M\"{o}nchhofstra\ss e 12-14, D-69120 Heidelberg, Germany
\\
$^{3}$ Special Astrophysical Observatory, Russian Academy of Sciences, Nizhnii Arkhyz 369167, Russia
\\
$^{4}$ Astronomical Institute of the Czech Academy of Sciences, Fri\v{c}ova 298, 25165 Ond\v{r}ejov, Czech Republic
}
\date{Accepted XXX. Received YYY; in original form ZZZ}

\pubyear{2020}

\begin{document}
\label{firstpage}
\pagerange{\pageref{firstpage}--\pageref{lastpage}}
\maketitle

\begin{abstract}

We report the identification of the unusual emission-line stellar-like object in the nearby low-metallicity ($Z \sim 0.1 \mathrm{Z_{\astrosun}}$) dwarf galaxy NGC\,4068. Our observations performed with long-slit spectrograph and Fabry-Perot interferometer demonstrate high velocity dispersion in \Ha line, presence of \HeII$\lambda$4686\AA\ line and peculiarly low [S~\textsc{ii}]/[N~\textsc{ii}] fluxes ratio for this object. From observational data, we derived that the object represent a single star of high bolometric luminosity ($L_* \sim 1.5\times10^6 L_\odot$) surrounded by an expanding nebula with kinematical age of $t\sim0.5$~Myr. The nebula exhibits significant nitrogen overabundance ($\log({\rm N/O}) \sim -0.05$, that is by $\sim1.4$~dex higher than expected for low-metallicity galaxies). We suggested that this is a massive blue supergiant (BSG) or Wolf-Rayet (WR) star surrounded by its ejecta interacting with the interstellar medium. We calculated the models of the nebula using \textsc{cloudy} photoionization code, applying \textsc{cmfgen}-modelled BSG and WR stars as ionisation sources. We found a best agreement between the modelled and observed spectra for the model assuming ionization by low-metallicity WR star of mass $M_*\approx80\,\Msun$, ionizing the nebula through the strong wind and enriching the interstellar medium with nitrogen.

\end{abstract}




\section{Introduction}\label{sec:introduction}

Massive stars ($M>8M_\odot$) play a crucial role in shaping the interstellar medium (ISM) of the galaxies, regulating star-forming activity and highly impacting their evolution. 
Through three main channels of the stellar feedback (ionizing radiation, winds and supernovae (SNe) explosions) massive stars inject the energy and momentum into the  ISM \citep[e.g.][]{Krumholz2014}. Most massive stars and, in particular, luminous blue variables (LBVs) and Wolf--Rayet (WR) stars are especially important contributors to the cumulative pre-supernovae stellar feedback due to their strong winds and hard ionizing radiation  \citep{Crowther2007, 2020Galax...8...20W}, that is indeed observed in the extragalactic \HII regions \citep[e.g.][]{Afanasiev2000, Egorov2018, Ramachandran2019, Bestenlehner2020}. 

Massive stars are key elements in the baryon cycle in the Universe. Heavy elements build up in nucleosynthesis during their evolution, and then they are injected into the ISM via stellar winds and supernovae. However, the direct observations of the local enrichment of the ISM by newly produced heavy elements are complicated because both global dynamical processes and small-scale turbulence driven by stellar feedback mix the metals in ISM at relatively short timescale ($\sim100$~Myr, \citealt{KrumholzTing2018}). Indeed, the measurements of the metallicity variations beyond the radial gradients demonstrate that the ISM is highly homogeneous at the scales of hundred parsecs \citep{Kreckel2020, Williams2022}. The attempts made to reveal the local signs of the enrichment of the ISM in the vicinity of massive stars (especially WR stars) also demonstrated normal (or even deficient) oxygen abundance (a common tracer of gas phase metallicity) in extragalactic \HII regions \citep[e.g.][]{ LS2007, LS2011, Maryeva2018, Kumari2018} as well as in the resolved nebulae in our Galaxy \citep[e.g.][]{Esteban1990, PM2013, Esteban2016}. At the same time, all these studies reveal the enhancement of the N/O ratio towards the WR stars, pointing to their importance in the enrichment of the ISM by secondary nitrogen in the galaxies \cite[e.g.][]{Roy2021}. 

\begin{figure*}
		\begin{minipage}[h]{0.99\linewidth}
			\center{\includegraphics[width=17cm]{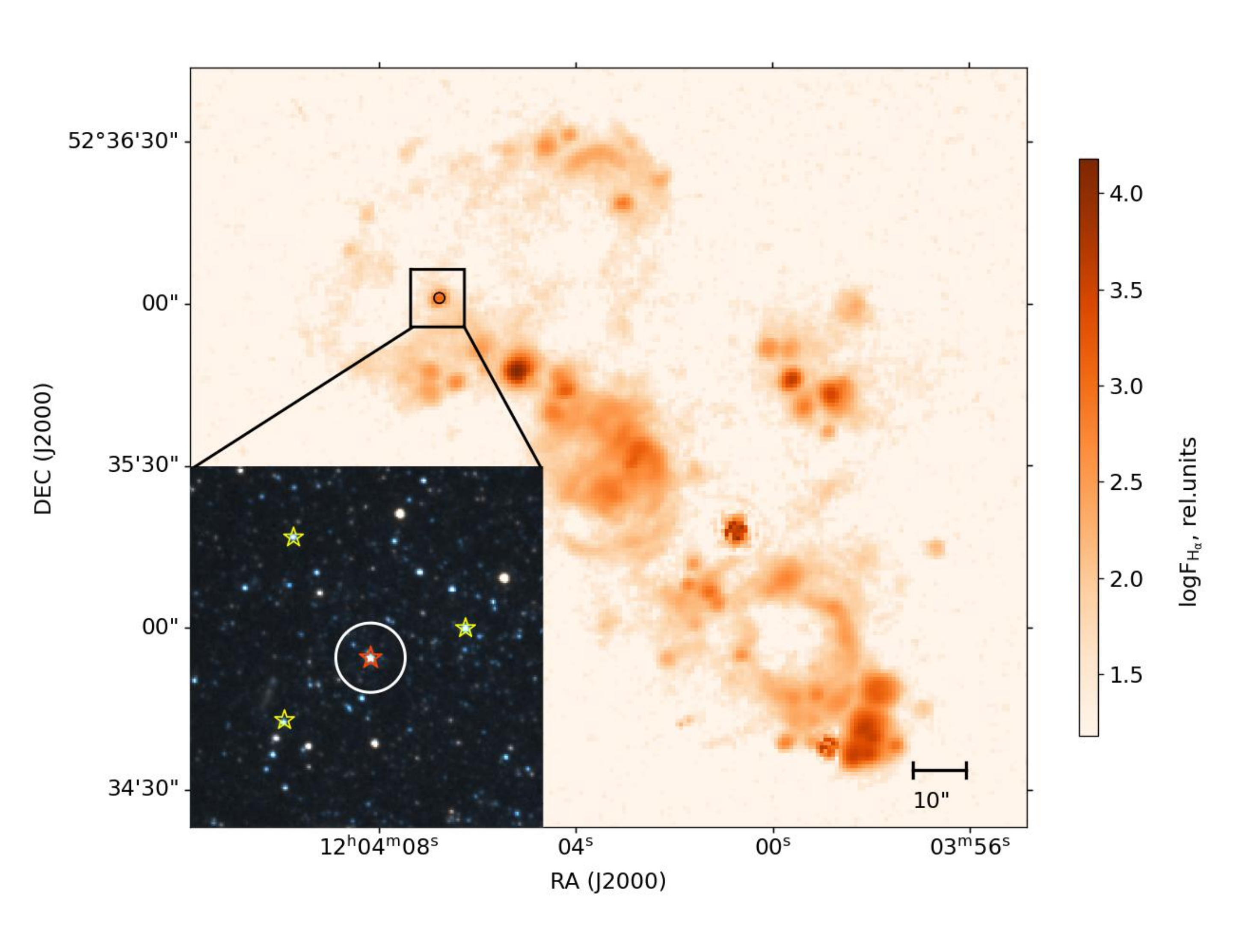}}
		\end{minipage}
		\caption{Map of NGC\,4068 in \Ha line (from our FPI observations), and imprinted {\it HST} image of the selected area around Object\,\#A. \Ha-map: position of Object\,\#A is marked by black circle, also corresponding to the PSF of the image (PSF = 2").  {\it HST} image $F606W$ and $F814W$ filters: red star marks the Object\,\#A, yellow stars mark other nearby O-type stars candidates, located in the studied region of the galaxy (see Sec.~\ref{est_sourse}). The white circle marks the area corresponding to the PSF-region of the \Ha map. Note that only one {\it HST}-resolved star is detected within the assumed borders of the nebula surrounding the Object\,\#A. }
		\label{Ha_Hubble_map}
\end{figure*}

Cumulative effects of feedback from massive stars at low metallicity are still not well understood. Both stellar winds \citep{Vink2001, Vink2021} and the rate of SNe type II \citep{Anderson2016} decrease with metallicity. Evolutionary tracks of massive stars also differ from those in our Galaxy \citep[e.g.][]{Stromlo2021}. 
Meanwhile, the impact of the stellar feedback to the small-scale kinematics of the ISM is clearly observed in the nearby low-metallicity dwarf galaxies \citep[e.g.][Gerasimov et al., in prep.]{Moiseev2012, Pustilnik2017, Egorov2017, Egorov2021}, though it is usually not easy to evaluate the relative impact of pre-supernovae feedback. Analysing the individual massive stars in low-metallicity environment and how they interact with the surrounding ISM, one can shed the light onto the physics of stellar feedback in such an extreme environment, resembling the conditions in the early Universe. Current sample of the well-studied massive stars at the metallicities below that of Small Magellanic Cloud is however still scarce \cite[see, e.g.,][for a review]{Garcia2021}.

In the current paper, we report the identification of the peculiar isolated ionized nebula in the nearby low-metallicity dwarf galaxy NGC\,4068 (Fig.~\ref{Ha_Hubble_map}) and argue that it is associated with a single massive WR star. The nearby dwarf irregular galaxy NGC\,4068 is located at the distance $D\sim4.36$~Mpc \citep{Makarov2013} in the Canis Venatici\,I cloud. Its gas-phase metallicity estimates has been never published in literature, but our long-slit spectral observations are indicative of the effective (at 0.4$R_{25}$) oxygen abundance $\mathrm{12+\log(O/H) \sim 7.85}$ ($Z\sim 0.15 Z_\odot$) reaching $\mathrm{12+\log(O/H) \sim 7.4}$ ($Z\sim 0.05 Z_\odot$) at the periphery (the analysis will be presented in subsequent paper Moiseev et al., in prep.) This estimate is in agreement with the luminosity of NGC\,4068 ($M_B \sim -15.1$ \citealt{Kaisina2012}) corresponding to $\mathrm{12+\log(O/H)} \sim 7.8$ according to the luminosity--metallicity relation from \citet{Berg2012}. 

\begin{figure*}
		\begin{minipage}[h]{0.49\linewidth}
			\center{\includegraphics[width=8.5cm]{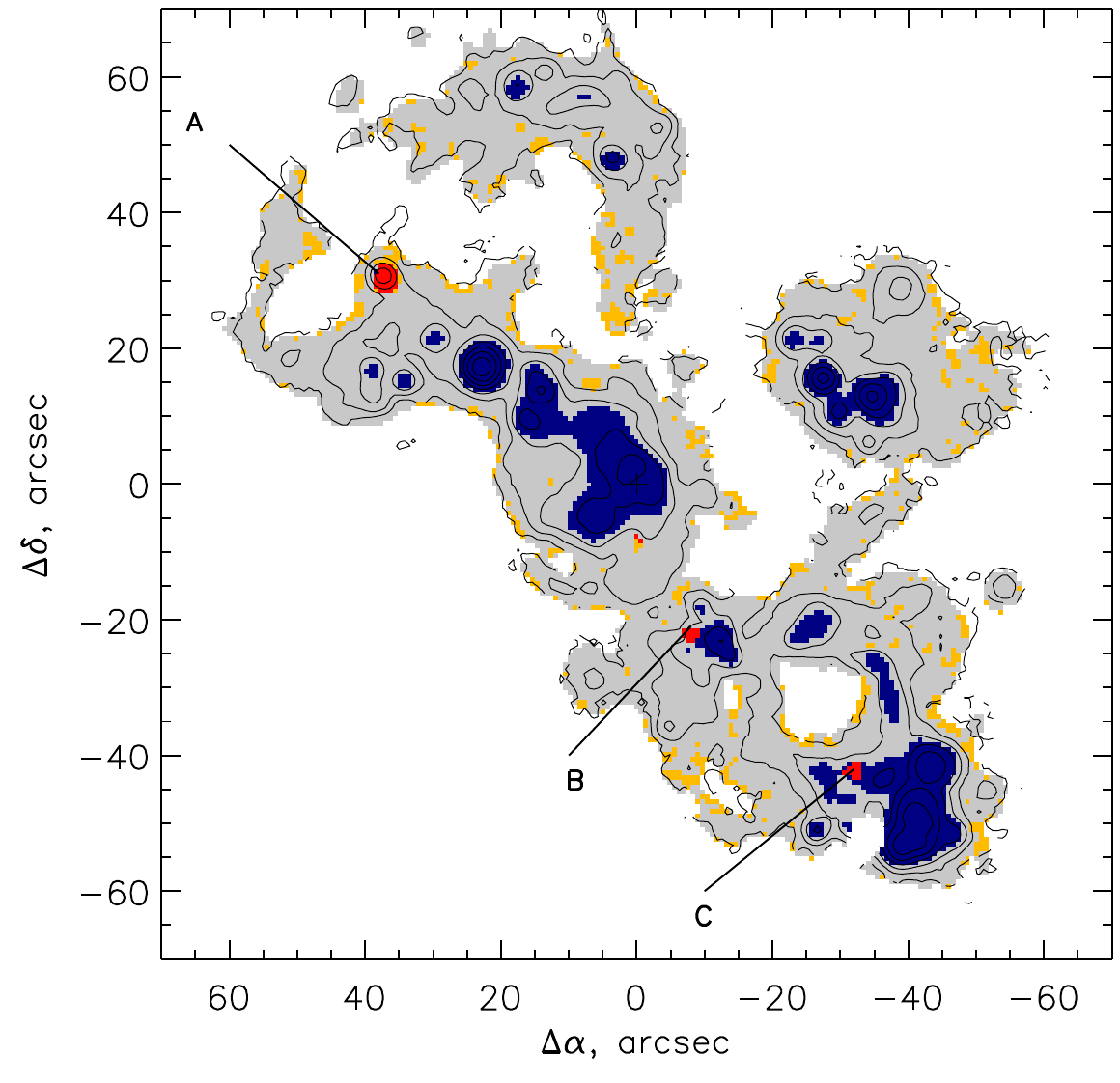} }
		\end{minipage}
		\hfill
		\begin{minipage}[h]{0.49\linewidth}
			\center{\includegraphics[width=9cm]{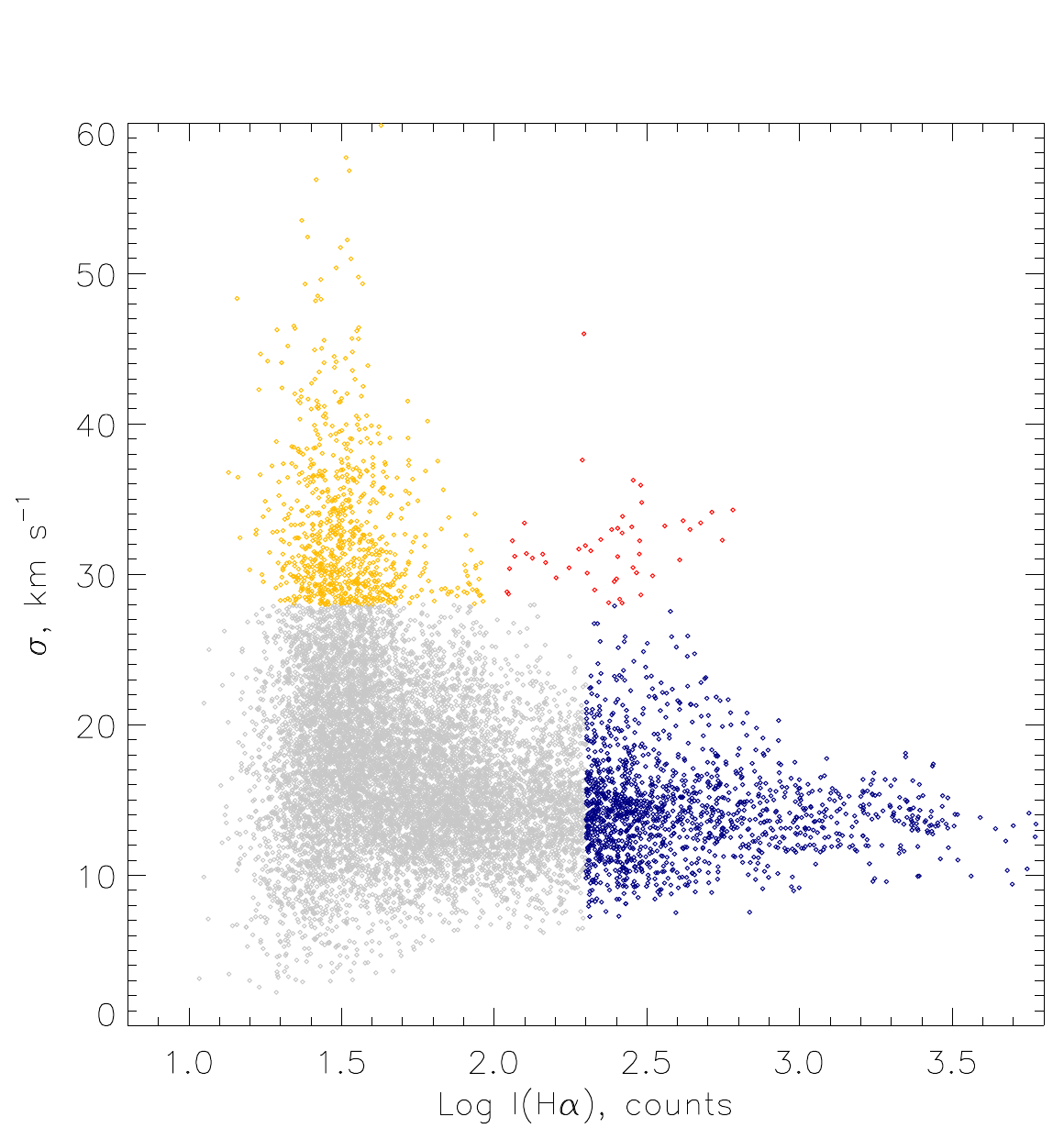} }
		\end{minipage}
		\caption{Right-hand panel:  NGC\,4068 map in \Ha line. Left-hand panel: I-$\sigma$ diagram of the galaxy. The colours of the points on the diagram correspond to the colour that indicate the particular areas on the galaxy map. Orange and grey colour mark the areas of diffuse gas, blue - the \HII-regions. Objects \#A, \#B, \#C lie on the diagonal of the diagram (red colour), which implies that they could correspond to expanding nebulae formed by strong stellar winds.
		}
		\label{isigma}
\end{figure*}

Analyzing the internal kinematics of ionized gas in the galaxy NGC\,4068 we found three compact nebulae having high both surface brightness and line of sight velocity dispersion in \Ha line (they are denoted as \#A, \#B and \#C in Fig.~\ref{isigma}). As was previously shown by \cite{Moiseev2012}, combination of such characteristics is typical for objects like WR, LBV, supernova remnants (SNR) and other energetic stellar sources. Subsequent long-slit spectroscopy demonstrated that while the objects \#B and \#C have emission lines fluxes ratio typical for SNRs, the Object\,\#A (RA = 12h04m06.8s, Dec = $52\degr36\arcmin00\arcsec$) exhibits very unusual spectrum with peculiarly faint \SII $\lambda 6717,6731$ ~\AA\, lines and highest ratio of \NIIHa\, in the galaxy (Fig.~\ref{spectrum_SAO},\ref{spectrum_CMO}). While the general analysis of the ionized gas in NGC\,4068 will be presented in a forthcoming paper by Moiseev et al. (in prep.), this paper is focused on the analysis of the peculiar Object\,\#A. We performed the modelling of its spectrum with the \textsc{cmfgen} and \Cloudy{} codes, aimed to reveal the nature of this object.

The paper is organized as follows. In Section~\ref{sec:observations} we describe the spectroscopic and photometric observations been performed, and in Section~\ref{sec:prop} we estimate the properties of the Object\,\#A based on the observations. We examine the possible nature of the Object\,\#A in Section~\ref{sec:nature}, and describe our modelling of the source of ionization and surrounding nebula with \textsc{cmfgen} and \Cloudy{} in Section~\ref{sec:models}. In Section~\ref{sec:discussion} we discuss the obtained results, which are summarized in Section~\ref{sec:summary}.

\begin{table*}
	\caption{Log of observational data: name of data set, date of the observation, exposure time $T_{exp}$, field of view (FOV), spatial resolution (Sp. res),  seeing ($\theta$), spectral range, spectral resolution ($\delta\lambda$) and dispersion ($\delta$) }
	\label{param_t}
	\begin{tabular}{llclllllll}
	\hline
	Data set              & Date     & $T_{exp}$, s & FOV                             & Sp. res., $''/pix$             & $\theta$, $''$   & Sp. range, \AA                 & $\delta\lambda$, \AA     & $\delta$, \AA/pix   \\ 
	\hline
	\multicolumn{9}{c}{Long-slit spectroscopy}\\
	SCORPIO-2/BTA PA=$43\degr$ & 2014 Feb 25 & 2700 & {$1\arcsec\times6.1\arcmin$}  & 0.36 &  1.0   & 3600-7070  & 4.8 &  0.87 \\
    SCORPIO-2/BTA PA=$51\degr$ & 2015 Apr 24 & 2700 & {$1\arcsec\times6.1\arcmin$}  & 0.36 &   1.8   & 3600-7070  & 4.8 &   0.87\\
    \multirow{2}{*}{TDS/CMO PA=$41\degr$} & \multirow{2}{*}{2020 Feb 23} & \multirow{2}{*}{$9\times600$} & \multirow{2}{*}{{$1\arcsec\times3\arcmin$}} & 0.35 & \multirow{2}{*}{1.5} & 3600-5770   & 2.4& 1.21\\
                                      &                               &                               &                   &        0.37         &                  & 5670-7460  &  2.6&0.87\\
                        
    \multirow{2}{*}{TDS/CMO PA=$131\degr$} & \multirow{2}{*}{2020 Dec 25,27} & \multirow{2}{*}{$8\times1200$} & \multirow{2}{*}{{$1\arcsec\times3\arcmin$}} & 0.35 & \multirow{2}{*}{1.7} & 3600-5770   & 2.4 &1.21\\
                                      &                               &                               &                   &       0.37          &               & 5670-7460   & 2.6 & 0.87 \\
\\                                      
    \multicolumn{9}{c}{Direct images and Fabry--Perot spectroscopy at the 6m telescope with SCORPIO-2} \\    
    Image AC5014  & 2021 May 06 & 1800 & $5.8'\times4.8'$ & 0.4 &     1.6         & [OIII] 5007 \AA  & 32 & \\

    FPI  & 2012 Mar 18 & 6400 & $6.1'\times6.1'$ &  0.7   &      1.2        &  \Ha & 0.4 & \\
	\hline
	\end{tabular}
\end{table*}

\section{Observations}\label{sec:observations}

Observations of Object\,\#A were carried out at two telescopes: 
6-m  Big Telescope Alt-azimuthal (BTA) of Special Astrophysical Observatory of Russian Academy of Science (SAO RAS) and 2.5-m telescope of Caucasian Mountain Observatory (CMO) of Moscow State  University. 
The spectra from CMO were obtained with the Transient Double-beam Spectrograph    \citep[TDS, ][]{2020AstL...46..836P}, while SCORPIO-2  multimode focal reducer \citep{scorpio2} was used for  observations in SAO RAS. The joint observational log is given in Table~\ref{param_t}. 

\subsection{Long-slit spectroscopic observations}\label{longslit_spectroscopic_obs}

The first two spectra of Object\,\#A were obtained with the SCORPIO-2 in 2014 -- 2015 in the spectral range 3600-7070~\AA\ with   6~arcmin long slit with a width of 1~arcsec. 
The spectral resolution $(\delta\lambda)$ estimated as the full width at half maximum (FWHM) of air-glow emission lines was 4.8~\AA.
The data reduction was performed in a standard way using {\sc IDL}-based pipeline, as described in our previous papers \citep[e.g.][]{Egorov2018}.

\begin{figure}
		\centering
		\includegraphics[width=1\linewidth]{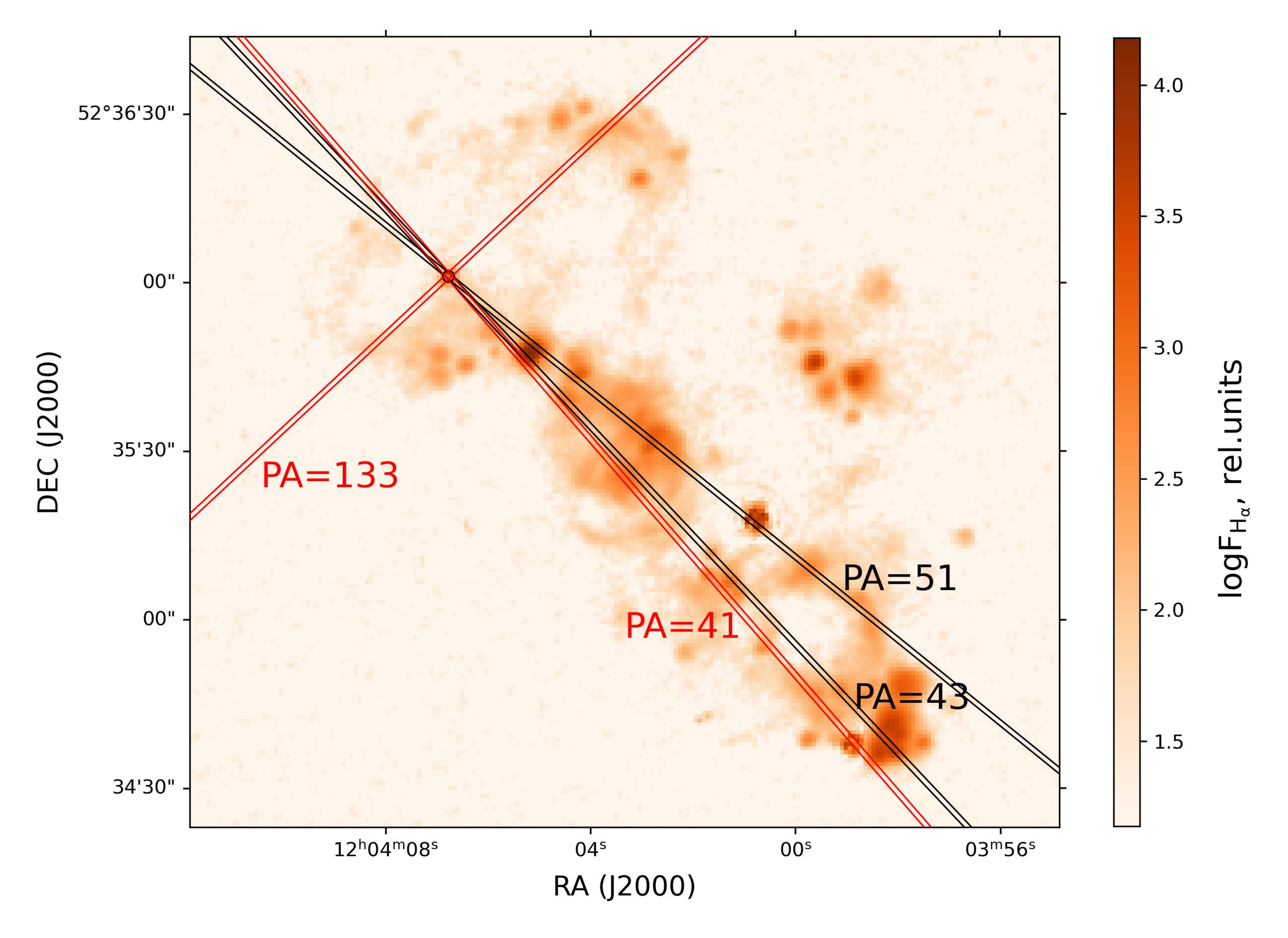}
		\caption{Slit positions of the TDS and SCORPIO-2 spectra overlaid on the \Ha image. Red lines correspond to TDS spectra, black -- to SCORPIO-2 spectra.}
	\label{slitpos}
	\end{figure}

The other two spectra were obtained with the TDS  in 2020 simultaneously in the green (3600-5770~\AA) and red  (5670-7460~\AA) arms. The slit length was 3~arcmin and its width was 1~arcsec. The spectral resolution, estimated in the same way as for SCORPIO-2 data, is given in Table~\ref{param_t}. The data reduction was performed using our  {\sc python}-based  pipeline with the same data reduction steps as for SCORPIO-2 spectra. All the TDS spectra were obtained during the commissioning period.
The location of the spectrograph slits on the galaxy image for both devices is shown in Fig.~\ref{slitpos}.

We extracted the 1D-spectra in the apertures corresponding to 2--3 of the seeing size visually verifying that the Object~\#A lies entirely in the aperture. To measure the fluxes of emission lines,  we used our own software working in {\sc python}. We applied Gaussian fitting to measure the integrated line fluxes of the studied region. To estimate the fluxes of the faint emission lines \SII$\lambda$6716, 6731, \OIII$\lambda$4363 and \HeII$\lambda$4686 we fixed their velocities to these of the bright lines of the similar ionization state (\NII$\lambda6584$ or \OIII$\lambda5007$). The uncertainties of the measured line fluxes were propagated through all the data reduction steps. The reported lines fluxes ratios were corrected for reddening, assuming the colour excess $E(B-V)=0.11^m$ as derived from the observed Balmer decrement (based on the stacked SCORPIO-2 spectra). For reddening correction, we utilized the \cite{Cardelli1989} curve parametrized by \cite{Fitzpatrick1999}. Flux ratios of the most important emission lines and their uncertainties are given in Table~\ref{tab:rel_fluxes}. 

\begin{figure*}
		\begin{minipage}[h]{0.99\linewidth}
			\center{\includegraphics[width=17cm]{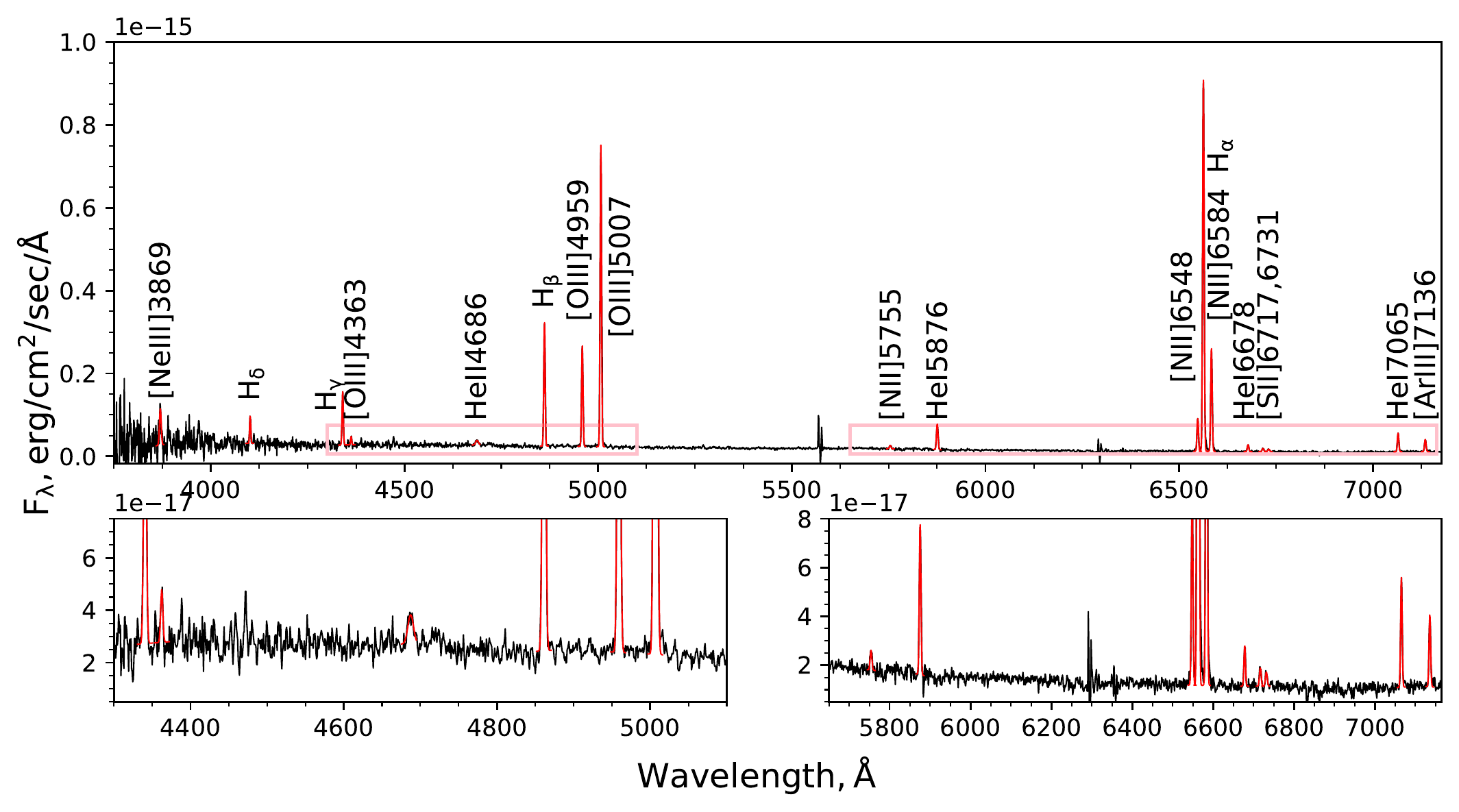}}
		\end{minipage}
		\caption{Combined SCORPIO-2/BTA spectrum (showed in black). The red line shows the results of performed Gaussian fitting of emission lines. The spectrum shows unusually low [S~\textsc{ii}]/\NII ratio and presence of \HeII $\lambda 4686$ line (bottom panels show these features in more detail).
		}
		\label{spectrum_SAO}
\end{figure*}

\begin{figure*}
		\begin{minipage}[h]{0.99\linewidth}
			\center{\includegraphics[width=17cm]{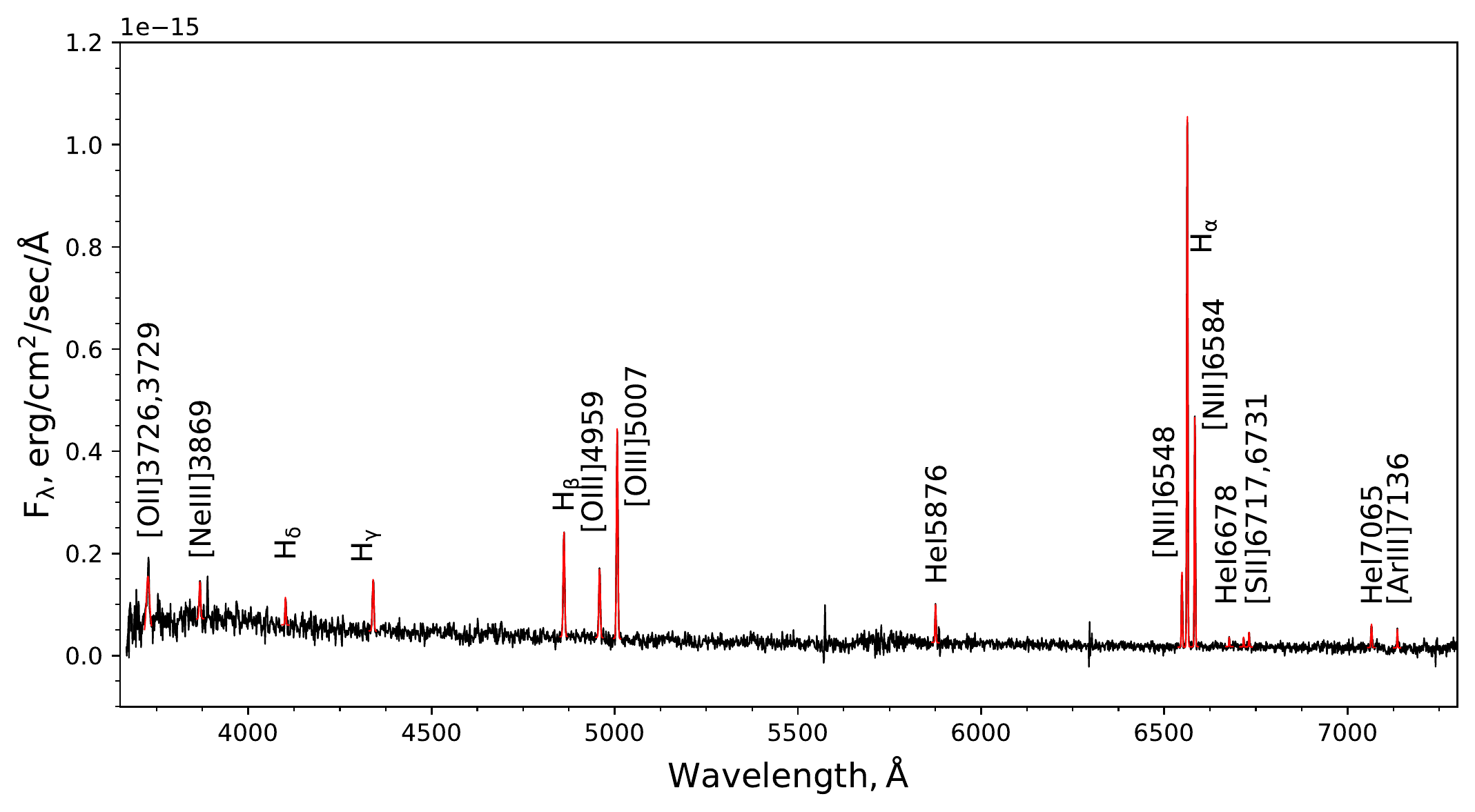}}
		\end{minipage}
		\caption{TDS/CMO 2.5-m telescope spectrum (PA=41\degr). Colors mark the same as in Figure \ref{spectrum_SAO}. \HeII $\lambda 4686$ line is not visible in the spectrum due to the lower S/N ratio.
		}
		\label{spectrum_CMO}
\end{figure*}

In all data sets, we used the  spectra of  the spectrophotometric standard stars obtained at a close zenith distance immediately after or before the object to calibrate the NGC\,4068 spectra to the absolute intensity scale.  
The sum of the two SCORPIO-2 spectra  is shown in Figure~\ref{spectrum_SAO}, while Figure~\ref{spectrum_CMO} presents the spectrum of Object\,\#A obtained with TDS in February 2020. In the spectra we see emission lines including bright \OIII $\lambda4959,5007$ and a very high ratio of [N~\textsc{ii}]/\SII unusual for \HII regions; in the TDS spectra, emission lines in the blue part of the spectra are also clearly detected. 

The \HeII$\lambda$4686 emission line is detected in the first obtained SCORPIO-2 spectrum (along PA=$43\degr$). 
Its equivalent width in stacked SCORPIO-2 spectra is $EW=4.5$~\AA, the fluxes ratio He~\textsc{ii}/\Hb$=0.07 \pm 0.02$ (Table~\ref{tab:rel_fluxes}), $FWHM = 7 \pm 2$ \AA, signal-to-noise ratio $S/N \sim 7 $. The line is not clearly seen on the second SCORPIO-2 spectrum, probably, because of the lower seeing. The achieved signal-to-noise ratio (S/N) in TDS spectra was also insufficient. To check this, we assumed the same He~\textsc{ii}/\Hb fluxes ratio in all the spectra and compared the expected flux of \HeII line with the associated uncertainty 
$$\sigma_l = 3\sigma_c\sqrt{N+EW/\Delta},$$
where $\sigma_c$ represents the standard deviation in the nearby continuum, N is the number of pixels of FWHM of the line, EW is the equivalent width of the line and $\Delta$ is the dispersion of the spectrograph (in \AA ~per pixel). Indeed, for both TDS spectra and the second SCORPIO-2 spectrum the noise level is comparable to or exceeds the expected \HeII$\lambda$4686 line flux, which explains why this line has not been detected in these data.   

We also note a difference in the \SII $\lambda 6731$~\AA\ line flux in the SCORPIO-2 and TDS spectra in Table~\ref{tab:rel_fluxes}. The ratio of \SII $\lambda$6717/6731 is a tracer of electron density ($n_e$), and thus a significant change in this ratio can reflect the spatial variations of the $n_e$ in the nebula. On the other hand, the difference  in the \SII $\lambda 6731$~\AA\ flux is within the 1$\sigma$ uncertainties and thus can be caused by lower S/N in TDS spectra. 



\begin{table}
\caption{Reddening-corrected emission lines fluxes ratios measured in two spectra of the nebula surrounding the object~\#A. The reddening correction was performed using Balmer decrement derived from the SCORPIO-2/BTA spectrum}
\label{tab:rel_fluxes}
\begin{tabular}{lcccc}
\toprule
Line                     & $\mathrm{F_{CMO~(PA=41\degr)}} $  & $\mathrm{F_{BTA~(combined)}}$ \\
\midrule

\OII  $\lambda3727$/\Hb &     1.3 $\pm$ 0.2  &       $-$                           \\
\NeIII $\lambda3869$/\Hb  &     0.39 $\pm$ 0.09  &    0.5 $\pm$  0.1             \\
\HeII  $\lambda4686$/\Hb &    $<0.12$         &    0.07 $\pm$ 0.02       \\
\OIII $\lambda4363$/\Hb &         $-$     & 0.05 $\pm$ 0.03 \\
\OIII $\lambda4959$/\Hb &      0.63   $\pm$ 0.05         & 0.82 $\pm$ 0.01 \\
\OIII $\lambda5007$/\Hb &     1.83  $\pm$    0.09     &     2.45 $\pm$ 0.03        \\
\NII $\lambda5755$/\Hb  &      $-$      & 0.034  $\pm$ 0.009        \\
\HeI $\lambda5876$/\Hb &   0.22 $\pm$ 0.03  &    0.231 $\pm$ 0.007      \\
\NII $\lambda6548$/\Hb  &  0.41 $\pm$ 0.03   &    0.225 $\pm$ 0.007      \\ 
\Ha $\lambda6563$/\Hb  &        2.99  $\pm$ 0.15      &     2.86 $\pm$ 0.03        \\
\NII $\lambda6583$/\Hb  &   1.25 $\pm$ 0.07   &    0.78 $\pm$ 0.01       \\ 
\HeI $\lambda6678$/\Hb &    0.036 $\pm$ 0.016    & 0.051   $\pm$  0.005    \\
\SII $\lambda6716$/\Hb  & 0.04 $\pm$ 0.02  &   0.041 $\pm$   0.008      \\ 
\SII $\lambda6731$/\Hb  &  0.06 $\pm$ 0.02 &    0.03 $\pm$ 0.01       \\  
\HeI $\lambda7065$/\Hb &    0.13 $\pm$ 0.03   & 0.133 $\pm$ 0.006         \\
\ArIII $\lambda7136$/\Hb  &   0.07 $\pm$ 0.02     &    0.086 $\pm$ 0.007    \\

\bottomrule
\end{tabular}
\end{table}
    
\subsection{Optical FPI-observations}\label{Optical_FPI_observations}

The integral-field spectral observations in the \Ha emission line were performed with the scanning Fabry-Perot interferometer (FPI) mounted inside the SCORPIO-2 \citep{Moiseev2021}. We analyzed the final data cube presented earlier by \citet*{MoiseevKlypin2015}. The data cube  contains 40 channels around the red-shifted \Ha line, other parameters are given in the Table~\ref{param_t}. We fitted the observed spectra  with  multi-component Voigt profile, that yields flux, line-of-sight velocity and velocity dispersion (corrected for instrumental broadening) for each component as an output \citep{2008AstBu..63..181M, 2015AstBu..70..494M}.

\subsection{Photometric data}\label{photometric_data}

Direct image  in the \OIII $\lambda5007$ emission line was obtained at the SCORPIO-2   with  the narrow-band  filter AC5014 (the central wavelength (CW) is 5018.0~\AA, FWHM$=$32~\AA). We used standard data reduction processes including bias subtraction, flat-fielding, correction for variations of atmospheric extinction and seeing. All different exposures were aligned and combined using sigma-clipping to remove the cosmic hits and artefacts.
We made the aperture photometry of the Object\,\#A in \OIII line: $$m_{5007} = -2.5 \mathrm{log} F_{5007} -13.74$$  (following \citealt{2002ApJ...577...31C}) and obtained ${m_{5007}}$ = $-5.42_{-0.39}^{+0.29}$ mag. For that we used only images in AC5014 filter, the continuum was not subtracted. 

In this paper we also use the archival {\it Hubble Space Telescope (HST)} images of the NGC\,4068 galaxy obtained with Advanced Camera for Surveys (ACS) and Wide Field Camera (WFC) detector with the wide-band filters $F606W$ and $F814W$ (Proposal ID 9771, PI: Karachentsev), and the results of their photometry from
\citet{2008MNRAS.384.1544S}.

We also inspected ultraviolet imaging data of the region acquired by \textit{Swift} Ultraviolet/Optical Telescope (UVOT) in UVW2 filter (CW = 2000 \AA), but strong and complex underlying ultraviolet emission in the region prevented us from reliably detecting the nebula and estimating its extent in the ultraviolet range.

\section{Properties of the Object \#A derived from observations}\label{sec:prop}
\subsection{Observational properties of the nebula}\label{est_neb}

The nebula is unresolved in our images, and its extent in \Ha line along the slit also doesn’t exceed the angular resolution of the spectra. Given the adopted distance to the galaxy NGC\,4068 ($D=4.36$~Mpc), we estimate the maximum size of the nebula of Object\,\#A as 30~pc.

\begin{figure}
		\centering
		\includegraphics[width=1\linewidth]{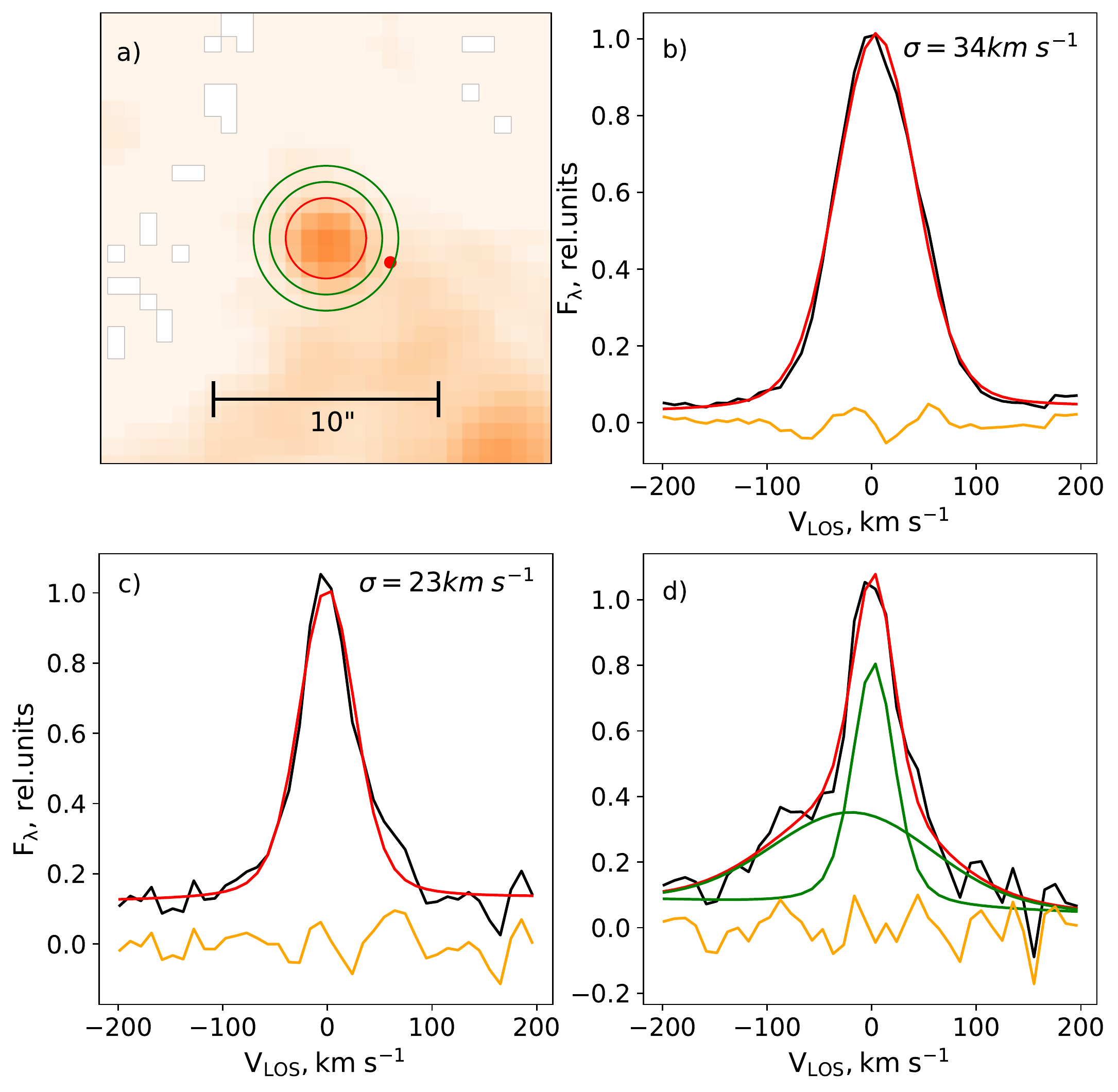}
		\caption{\Ha image of the vicinity of the object~\#A (panel a), and the examples of the \Ha line profile in FPI data. Panels (b) and (c) show the observed \Ha line profile (black color), results of its single Voigt component fitting (red color) and the residuals (yellow color) for the spectra integrated over the nebula (within the red circle on panel a) and over its surrounding gas (within the ring shown by green color on panel a). Panel (d) demonstrates the line profile towards the south-east of the nebula (exact position is shown by red dot on panel a), green color shows the individual components. A broad component can be produced by the interaction between the stellar wind and surrounding ISM.}
	\label{Ha_profile}
	\end{figure}

Initially, the object \#A has drawn our attention because of the high velocity dispersion in \Ha line in comparison with the other \HII regions in the galaxy (see Fig.~\ref{isigma}). We show the results of the Voigt fitting of the \Ha line profiles extracted integrated over the nebula and over its surrounding in Fig.~\ref{Ha_profile} (panels b and c, respectively). The wider \Ha line profile of the object \#A can be related with the expanding bubble around it, which can be driven by the powerful wind of the star. The line profiles show weak asymmetry, but cannot be decomposed onto the individual components corresponding to the approaching and receding sides of the bubble. A two-component line profile is, however, observed at the distance of $\sim 3$~\arcsec towards the south-east of the object~\#A (panel d) -- the broad component there can be associated with the emission behind the shock front in the region where the stellar wind impacts the surrounding ISM. 

Though we cannot directly measure the expansion velocity of the bubble surrounding the object~\#A, we can estimate it from the difference of the velocity dispersion using the method presented by \citep{2021AstBu..76..367S}. In the case of observations with IFP751, 
$$V_{\rm exp} = k(\sigma_{\rm obs}^2 - \sigma_{\rm ISM}^2)^a + v_0$$ 
where $\sigma_{\rm obs} = 34 \kms$ is the measured \Ha velocity dispersion in the object~\#A, $\sigma_{\rm ISM} = 23 \kms$ is the \Ha velocity dispersion of the surrounding ionized gas, and the coefficients $k\simeq3.2$, $a\simeq0.4$ and $v_0\simeq-3.3$ (note that these coefficients are dependent on $\sigma_{\rm ISM}$). Then, $V_{\rm exp}\simeq 38 \kms$. 


Given the obtained radius $R$ and expansion velocity $V_{exp}$, we can estimate the upper limit of the age of the nebula assuming the \cite{1977ApJ...218..377W} model describing the bubble evolution:
$$t = 0.6R/V_{\rm exp} \simeq 0.5\ {\rm Myr}.$$
The kinematic age of the bubble indicates the characteristic age from the moment of onset of the powerful wind. The obtained value is in agreement with the typical lifetimes of the massive stars at the WR or BSG stages.


\begin{figure}
    \centering
    \includegraphics[width=\linewidth]{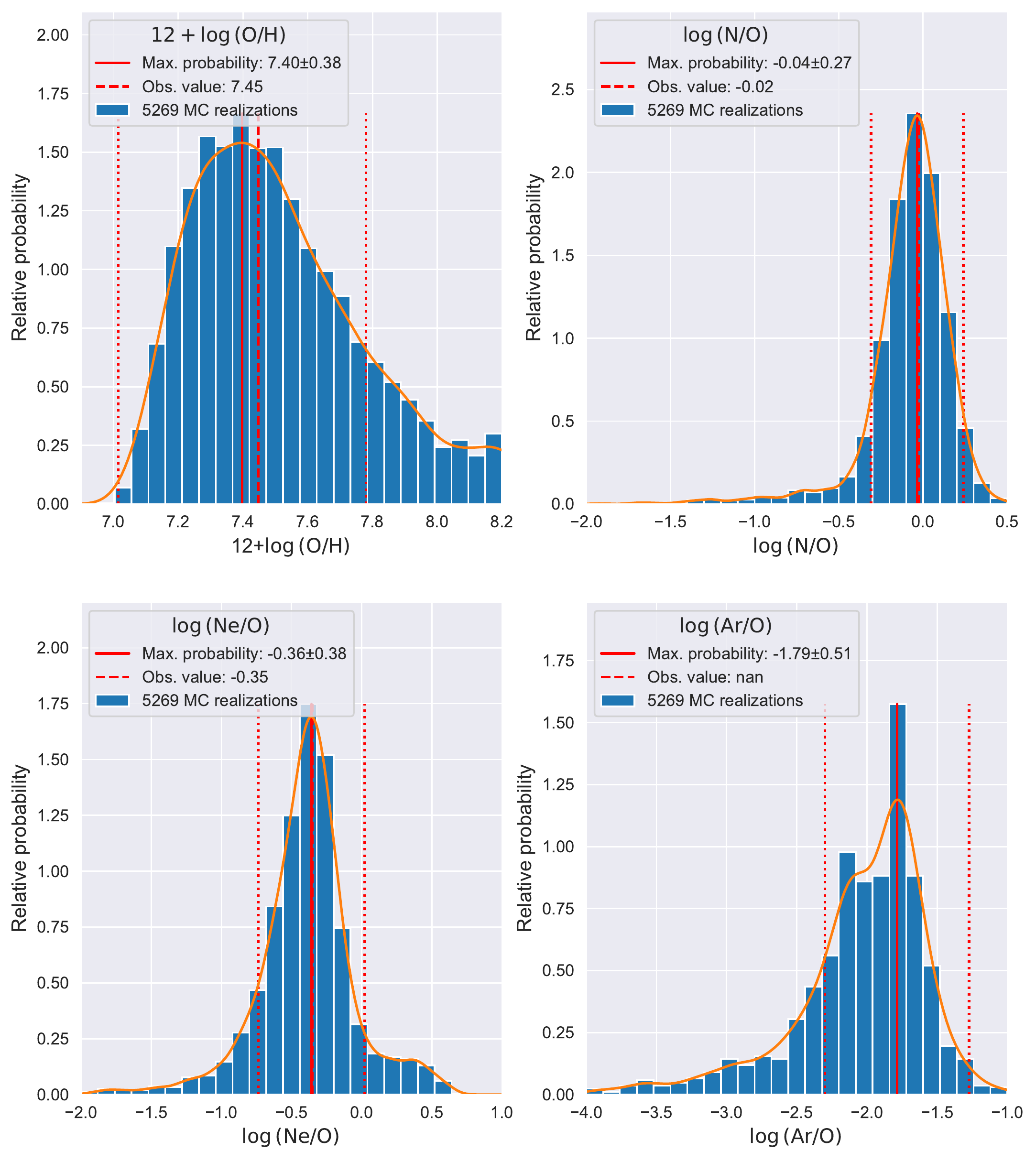}
    \caption{Chemical abundances of O/H, N/O, Ne/O, Ar/O measured for the nebula in Object~\#A with $T_e$ method. Histograms show the probability distribution of each value for a sample of $\sim 5300$ successful (with the line ratios valid for $T_e$ estimates) Monte Carlo realizations, where the fluxes of all emission lines were randomly distributed around the measured ones with a standard deviation equal to the observational uncertainties. The solid line shows the most probable value, dotted lines correspond to the standard deviation of the measured parameter. For reference, the dashed line shows the values obtained from the measured line fluxes.}
    \label{fig:abundances}
\end{figure}

The oxygen abundance (proxy of a gas-phase metallicity) in NGC~4068 is not uniform -- according to our long-slit SCORPIO-2 spectra, a remarkable gradient from ${\rm 12+\log(O/H)\sim7.8}$ in central regions to about 7.5 on the periphery is observed (Moiseev et al., in prep.). Relative nitrogen abundance in NGC~4068 is ${\rm \log(N/O) = -1.6 ... -1.3}$, typical for low metallicity dwarf galaxies (see, e.g., \citealt{1999cezh.conf..149I, 2006A&A...448..955I}). Based on these measurements and on the position of the object~\#A in the galaxy, we would expect it has ${\rm 12+\log(O/H) \sim 7.6}$ ($Z\simeq 0.08 Z_\odot$) and ${\rm \log(N/O) \sim -1.5}$ if it is a normal \HII region.

The faint sensitive to electron temperature $T_e$ auroral emission lines \OIII $\lambda4363$~\AA\, and \NII $\lambda5755$~\AA\, are well detected in SCORPIO-2 spectra (see Fig.~\ref{spectrum_SAO}), and thus we can estimate oxygen and nitrogen abundances using `direct' $T_e$-method. However, the \OII $\lambda3727,3729$~\AA\, lines are were not detected in SCORPIO-2 spectra because of the low sensitivity of the used detector at the blue wavelength. Therefore, we used the reddening-corrected values of [O~\textsc{ii}]/[N~\textsc{ii}] (given the similar ionization potentials of these ions) measured in the TDS spectra (Fig.~\ref{spectrum_CMO}). Using the \textsc{pyneb} package \citet{Luridiana2015}, we estimated the relative abundance of 4 elements: O/H, N/O, Ne/O and Ar/O based on the measured line ratios given in Table~\ref{tab:rel_fluxes} and using the ionization correction factors for N, Ne, Ar as defined in \cite{2006A&A...448..955I}.  To estimate the uncertainties of these values, we generated 7000 synthetic Monte Carlo observations with the fluxes of each emission line randomly distributed around the measured values. About 5300 of these realizations were appropriate for estimating $T_e$ in both low- and high-ionization zones -- $T_e$([N~\textsc{ii}]) and $T_e$([O~\textsc{iii}]) (see Fig.~\ref{fig:abundances}). As a result, we find that the object~\#A has low oxygen abundance ${\rm 12+\log(O/H)} =  7.40 \pm 0.38$ (in agreement with the metallicity distribution in NGC~4068) and significant nitrogen relative nitrogen abundance ${\rm \log(N/O) = -0.04 \pm 0.27}$ unusual for the low-metallicity galaxies. Object~\#A has an excess of nitrogen in the nebula by a factor of $20-30$ in comparison with standard values for the \HII regions of the same metallicity. The abundances of other analyzed elements is consistent within the uncertainties with those typically observed values in low-metallicity and WR-galaxies \citep{2006A&A...448..955I, LS2010}.

We also estimated the electron density \mbox{$n_e\sim120$~cm$^{-3}$} of the nebula in the object~\#A based from the ratio of [S~\textsc{ii}]6717\AA/[S~\textsc{ii}]6731\AA\, lines using the \textsc{pyneb} code. However, the low brightness of \SII lines make this estimate very uncertain.

\subsection{Observational properties of the central ionization source}\label{est_sourse}

Figure~\ref{Ha_Hubble_map} shows the \Ha map of the NGC\,4068 and the \HST images in $F606W$ and $F814W$ filters. The size of the point spread function (PSF) of \Ha map (which we consider as an upper limit of the nebula size) is shown on both \Ha and \HST maps by circles: there are no other OB stars that could ionize the nebula of Object\,\#A. To distinguish the OB stars we relied on the results of \HST photometry published in catalog by \citet{2008MNRAS.384.1544S} and corrected for the Galactic extinction assuming $A_{F606W} = 0.052$~mag and $A_{F814W} = 0.033$~mag \citep{2011ApJ...737..103S}. We applied the similar criteria as in \cite{Bastian2011}: $M_{V}< -5$ mag and $-0.5 < M_{V}-M_{I}< -0.1$ mag (which can be transformed to 
$m_{F606W}< -4.81$~mag and $-0.32<m_{F606W}-m_{F814W}<-0.08$~mag 
using the method from \citealt{2005PASP..117.1049S} and distance to NGC\,4068 $D=4.36$~Mpc). Object\,\#A is shown with a red star, and it is significantly (more than 1 mag) brighter in V band than the other three OB stars identified in the same region of the galaxy, but outside the optical borders of the nebula. 
This implies that the Object~\#A is the main source of ionization of the nebula. 

Object\,\#A has apparent magnitudes in $F606W$ and $F814W$ filters $m_{F606W}=21.145$~mag and $m_{F814W}=21.709$ mag, respectively, which correspond to the absolute magnitudes $M_{F606W} = -7.16$~mag and $M_{F814W} = -6.52$~mag after correction for the Galactic extinction. 
To estimate the magnitudes of the ionizing star free from the nebula lines contamination, we calculated the contribution from the emission lines in $F606W$ filter for the spectra shown in Figs.~\ref{spectrum_SAO}, \ref{spectrum_CMO} and found that it is equal to $0.24$ and $0.18$ mag, respectively. Therefore, we took an average value of $+0.21$ mag, by which we corrected the observed magnitude in $F606W$ filter ($M^*_{F606W} = -6.89$). Magnitude in the $F814W$ filter should not be affected significantly, as no bright nebular lines fall into the filter bandwidth. 
Finally, we transformed the magnitudes from $F606W$ and $F814W$ bands to Johnson-Cousins $V$ and $I$ bands following \citet{2005PASP..117.1049S}:  $M_V=-6.94$~mag and $M_I = -6.56$~mag. 

\begin{figure}
		\centering
		\includegraphics[width=1\linewidth]{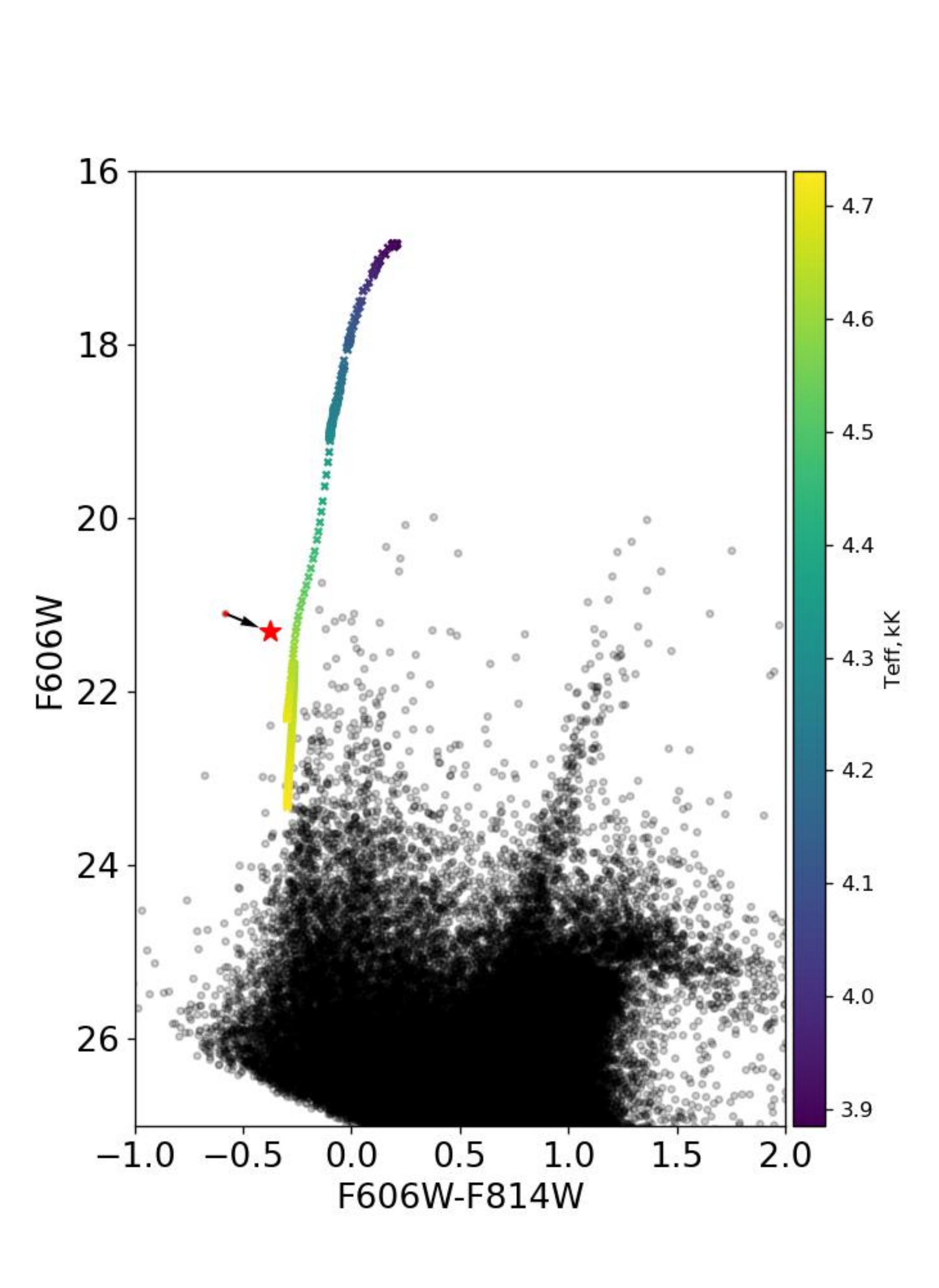}
		\caption{Color-magnitude diagram of the NGC\,4068 galaxy in {\it HST} $F606W$ and $F814W$ bands. The position of the Object\,\#A is marked by the red star on the diagram, whereas the red point corresponds to the values before correction for the contribution of the nebular emission lines. Geneva evolutionary track \citet{GenevaTrack} for a star with a mass of $80\Msun$ is overlaid, color marks the effective temperature of a star.}
		\label{HRdiagram1}
	\end{figure}

Usually, a color of a star allows one to roughly estimate its effective temperature. Object\,\#A  resides in the upper left corner of the color--magnitude diagram (Figure~\ref{HRdiagram1}), leftward from the Main Sequence in the region occupied by hot evolved stars.  Color $F606W-F814W$ is not sensitive to temperatures higher than $T_{\rm eff}>25$~kK, therefore we may only constrain the temperature of Object\,\#A to be $25<T_{\rm eff}<45$~kK. 
On the other hand, intensity of \OIII$\lambda5007$~\AA\, nebular emission line is dependent on the temperature of an ionizing source, and therefore \OIII~$\lambda5007$/\Hb ratio can be used to refine the temperature. We computed a grid of \Cloudy\, models of the nebula ionized by a blackbody source. We varied the effective temperature and luminosity of an ionizing source  and the electron density of a nebula in the ranges close to the  parameters of OB-type stars. Our calculations show that the observed value of \OIII~$\lambda5007$/\Hb ratio with the given parameters of the models cannot be obtained  if the temperature is less than $\log(T_{\rm eff})~=~4.6$ ($T_{\rm eff}\simeq40$~kK). This value of $T_{\rm eff}$ also corresponds to the closest position on the evolutionary track for a star of $80M_\odot$ to the object~\#A on the color-magnitude diagram (Figure~\ref{HRdiagram1}). We assume further this value as an effective temperature of the star in the Object~\#A.

From the measurements of the $M_V$ given above, we estimated the bolometric absolute magnitude and luminosity of the ionizing source in the object~\#A: ${M_{\rm bol}}=-10.74$~mag and $L_*=1.5\times10^6\,\Lsun$. For this, we applied the bolometric correction for the O-type stars (O3\,I), having effective temperature $T_{\rm eff}=42.2$~kK from \citet{MartinsPlez2006}. 

Note that for we corrected the measured magnitudes only for the Galactic extinction, but our long-slit spectra reveal higher value of extinction towards the Object~\#A (see Sec.~\ref{longslit_spectroscopic_obs}) remarkably exceeding the extinction in other \HII regions in NGC~4068 (Moiseev et al., in prep). This can be related, e.g., with localized dust production by the Object~\#A, and thus the total luminosity is probably underestimated. Given that usually stellar continuum is less affected by dust than gas ($E(B-V)_\mathrm{*} \simeq 0.44 E(B-V)_\mathrm{gas}$ according to \citealt{Calzetti2001}), corresponding values of extinction in the used \HST filters would be $A_{F606W} = 0.14$~mag and $A_{F814W} = 0.08$~mag. Usage of this value of extinction will almost  not affect the estimates of $V-I$ color presented above and slightly increase the bolometric luminosity by 0.14~mag leading to $L_*=1.7\times10^6\,\Lsun$.

\section{Possible nature of the Object \#A}\label{sec:nature}

Estimated luminosity of the Object\,\#A is very high, therefore we firstly checked whether it is a group of stars. {\it HST} images of the surrounding region demonstrate that  Object\,\#A is a single stellar-like object instead of a small cluster (Section~\ref{est_neb}). 
Observations of the object with TDS were carried out at different angles (see Sec.~\ref{longslit_spectroscopic_obs}), while the obtained spectra are identical, which also excludes the ionization of the nebula by cluster. 
Thus, we further consider the central star of Object\,\#A as a single object. 

The spectrum of the Object\,\#A looks very similar to that of some planetary nebulae (PNe). For example, the spectrum for one PN in the metal-poor galaxy Sextans\,B demonstrates the similar \SII and \NII emission line fluxes ratios as measured for the Object~\#A \citep{2005AJ....130.1558K}. 
However, in comparison with PNe, the Object\,\#A has very bright \OIII~$\lambda5007$ emission line. It exceeds the empirical upper limit of the \OIII~$\lambda5007$ flux established for PNe by an order of magnitude \citep[e.g.][]{2002ApJ...577...31C, Scheuermann2022}. Therefore, the Object\,\#A cannot be a PN, neither a more exotic type of PN with [WR] central star (see for example  \citealt{2022MNRAS.509..974G} and reference therein).

\begin{figure*}
	\centering
	\includegraphics[width=0.49\linewidth]{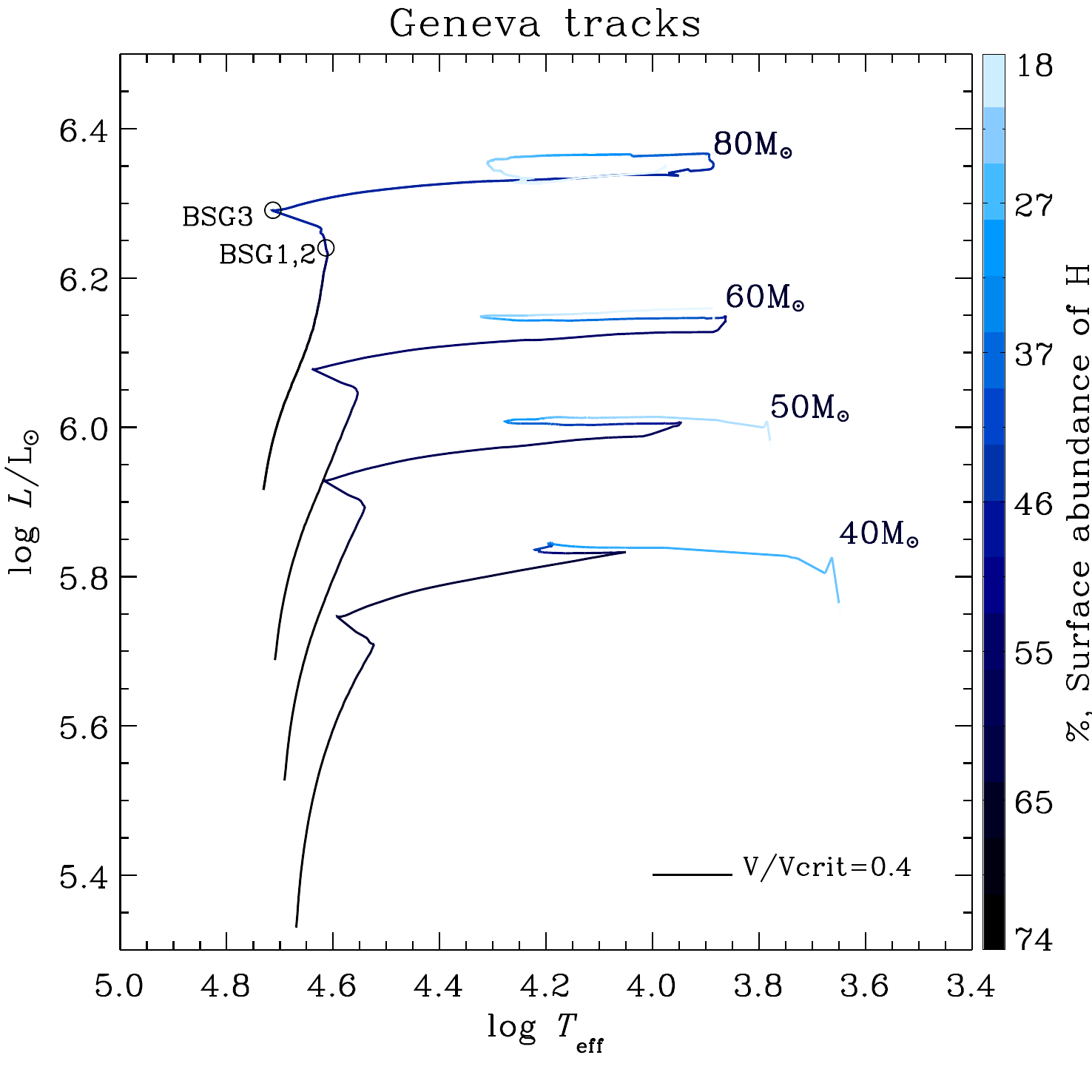}
	\includegraphics[width=0.49\linewidth]{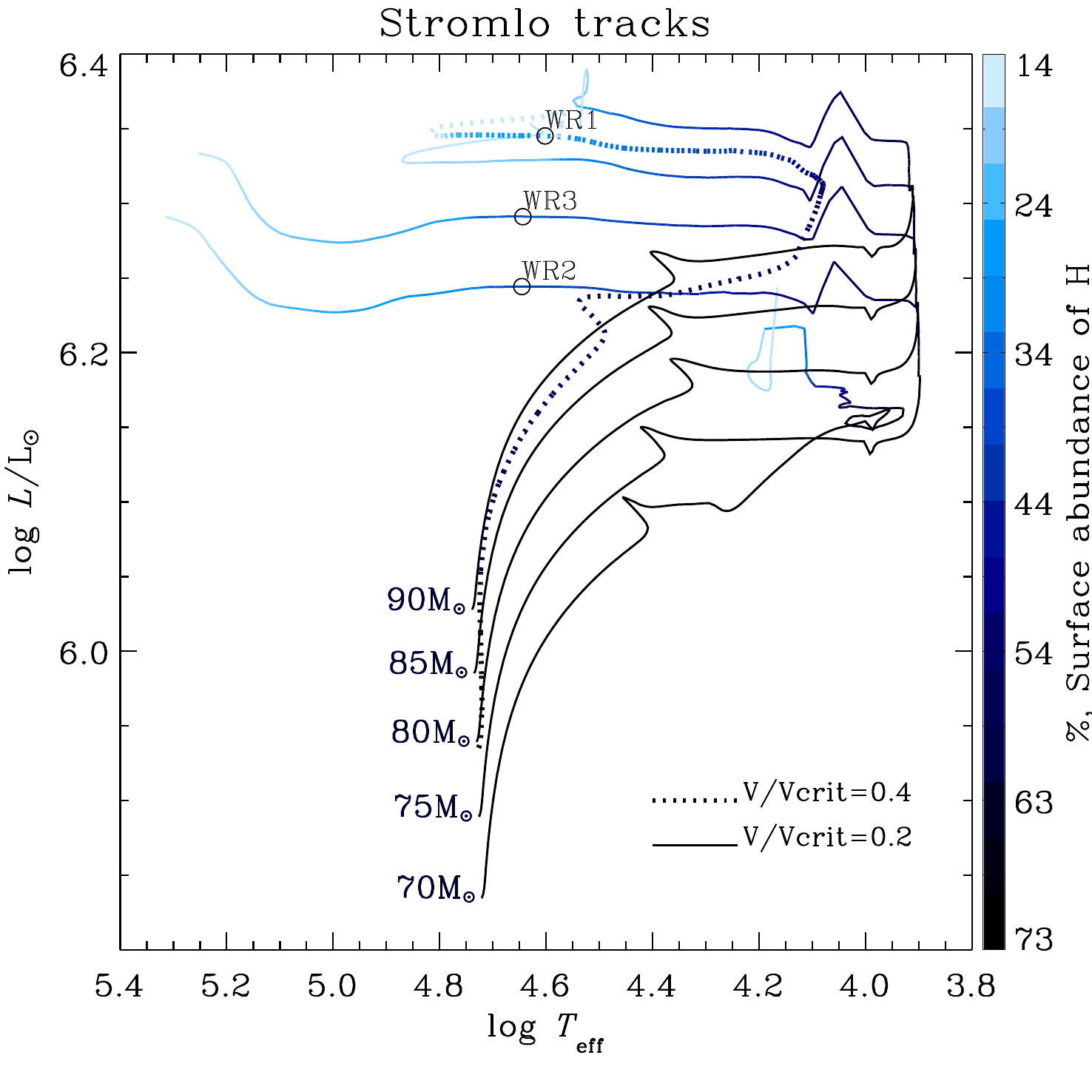}
	\caption{HR diagram and evolutionary tracks for the low metallicity massive stars (Z=0.002) from \citet{GenevaTrack} (left-hand panel) and \citet{Stromlo2021} (right-hand panel) evolutionary models.  
Black circles mark the points that we took as the ionization sources for further modelling. Geneva tracks do not produce massive WR stars at such low metallicities, while Stromlo tracks show rather different evolutionary path and produce WR stars with $M>70M_{\astrosun}$.}
\label{Tracks}
\end{figure*}

Taking into account the high luminosity of the object, its position in the color--magnitude diagram, strong \OIII~$\lambda5007$ line and significant  nitrogen overabundance of the nebula, we put forward a hypothesis about the ionization of the nebula by a massive star ($M_*=70-80\Msun$) at a late stage of evolution. Suitable objects are BSG,  LBV or WR stars. Estimated nebula parameters (size, $V_{\rm exp}$, luminosity) are consistent with the hypothesis of its ionization by a massive star. WR star is most likely to produce the observed ejection of nitrogen from its stellar atmosphere. Also, the estimated kinematical age of the expanding nebula around Object~\#A is consistent with the typical duration of the WR phase ($\sim0.4$~Myr according to \citealt{Groh2014}). 

\section{Modelling}\label{sec:models}

We model the Object\,\#A assuming it is a spherical nebula ionized by a massive single star with a powerful wind. In previous sections, we have shown that this scenario is consistent with observational data and estimated parameters of the emission nebula and the ionizing star. 
Modeling was performed in two steps. At the first step (Sec.~\ref{mod_of_st}) we simulated the spectrum of the ionizing star using the {\sc cmfgen} code \citep{Hillier5}. This spectrum was then used for the next step as an ionizing source for the \Cloudy{} model of the nebula (Sec.~\ref{mod_of_neb}).

\subsection{Modelling of central star}\label{mod_of_st}

As described in Section~\ref{est_sourse} we estimated the luminosity $L_*$ and effective temperature $T_{\rm eff}$ based on {\it HST} photometry and  color--magnitude diagrams. It allowed us to determine the position of Object\,\#A on Hertzsprung--Russell (HR) diagram (Figure~\ref{Tracks}). On the next step, we derived chemical abundances and physical properties of its central star from evolutionary tracks calculated by \citet{GenevaTrack} and \citet{Stromlo2021}, which differ in method of scaling abundances with total metallicity \citep{Nicholls2017}.

\begin{figure}
		\centering
		\includegraphics[width=1\linewidth]{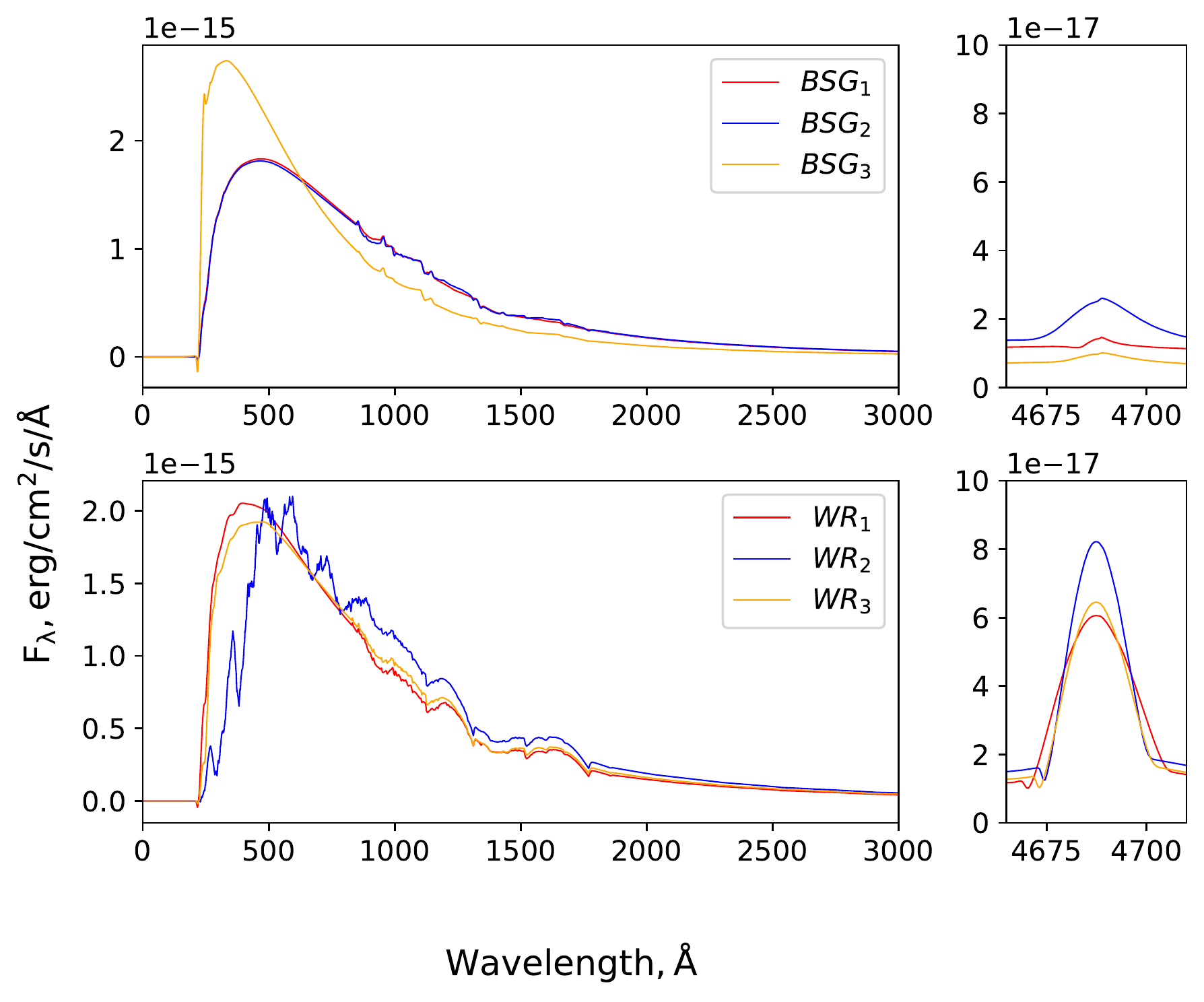}
		\caption{Left-hand panel: spectral energy distributions (SEDs) calculated for different models of central star. Right-hand panel: intensity of \HeII~$\lambda4686$ line in different models. Only WR stars have bright enough \HeII $\lambda4686$ line in their modelled spectra.}
		\label{SEDs}
	\end{figure}

\begin{table*}
\caption{Parameters of {\sc cmfgen} model for the central source.  
} 
\label{tab:cmfgenparameters}
\begin{tabular}{l lll || lll}
\toprule
                   &\multicolumn{3}{c||}{Geneva Tracks }                  &  \multicolumn{3}{c}{Stromlo Tracks}               \\
    & \multicolumn{1}{c}{Model} & \multicolumn{1}{c}{Model} & \multicolumn{1}{c||}{Model} & \multicolumn{1}{c}{Model} & \multicolumn{1}{c}{Model} &  \multicolumn{1}{c}{Model}  \\
    & \multicolumn{1}{c}{$\rm BSG_1$}   & \multicolumn{1}{c}{$\rm BSG_2$}  & \multicolumn{1}{c||}{$\rm BSG_3$} &  \multicolumn{1}{c}{$\rm WR_1$}  & \multicolumn{1}{c}{$\rm WR_2$} &  \multicolumn{1}{c}{$\rm WR_3$} \\
 \midrule
age (Myr)                    &  3.41            &   3.41             &     3.7             &  3.87            &  3.70              & 3.61  \\
$M_{\rm init}$  ($\Msun$)    &  80              & 80                 &     80              &  80              &  75                &  80   \\ 
$M_*$ ($\Msun$)              &  74.33           & 74.33              &     73.11           &  50.97           &  44.07             & 47.83 \\
${\rm log}\ g$               &  3.48            & 3.48               &    3.82             &  3.163           &  3.375             & 3.356  \\
$L_*$ ($10^6\,\Lsun$)       & 1.7               & 1.7                & 1.95                & 2.214            & 1.755              &  1.954  \\
${\rm log} \ (L_*/\Lsun)$    & 6.23             & 6.23               & 6.29                & 6.345            & 6.24               &  6.29  \\
${\dot{M} (\Msun~yr^{-1}})$  &$2.20\cdot10^{-5}$& $2.20\cdot10^{-5}$ & $2.708\cdot10^{-5}$ &$5.70\cdot10^{-5}$& $3.73\cdot10^{-5}$ & $4.46\cdot10^{-5}$  \\
${\rm log}{\ T_{\rm eff}}$   & 4.61             & 4.61               & 4.71                &  4.60            & 4.65               & 4.64  \\
$V/V_{\rm crit}$          & 0.4             & 0.4               & 0.4                &  0.4            & 0.2               & 0.2  \\
${V_\infty (\kms)}$          & 2760             & 2760               & 3360                &  1267            & 1147               & 1275  \\
$f$                          &1.0               & 0.1                & 0.5                 &  0.1             & 0.1                & 0.1  \\
\\
\midrule%
   & \multicolumn{6}{c}{Abundances (mass fraction)} \\
H\,centr                       &        9.6        &   9.6           &      0             &  0                        &    0                      & 0  \\
He\,centr                      &       90.2        &   90.2           &    99.8           & 3.8                       &    17.8                   & 16.1  \\
 ${\rm H_{surf}}$            &0.56               &0.56              & 0.48                &  0.29                     & 0.34                      & 0.32     \\ 
 ${\rm He_{surf}}$           &0.43               &0.43              & 0.51                &  0.70                     & 0.65                      & 0.67    \\
 ${\rm 12C_{surf}}$          &$1.23\cdot10^{-4}$ &$1.23\cdot10^{-4}$& $9.88\cdot10^{-5}$  & $3.44\cdot10^{-5}$        & $5.28\cdot10^{-5}$        &  $4.48\cdot10^{-5}$   \\ 
 ${\rm 14N_{surf}}$          &$7.59\cdot10^{-4}$ &$7.59\cdot10^{-4}$& $8.44\cdot10^{-4}$  & $2.86\cdot10^{-3}$        & $2.71\cdot10^{-3}$        &  $2.77\cdot10^{-3}$   \\
 ${\rm 16O_{surf}}$          &$3.25\cdot10^{-4}$ &$3.25\cdot10^{-4}$& $2.60\cdot10^{-4}$  & $5.22\cdot10^{-5}$        & $2.06\cdot10^{-4}$        &  $1.48\cdot10^{-4}$   \\  
\bottomrule 
\end{tabular}
\end{table*}
\citet{GenevaTrack} calculated evolutionary tracks for stars in a wide range of initial masses at low metallicity. According to their calculations, metal-poor stars become colder  after the end of hydrogen burning in the core and move along the track to the right side of the diagram (left panel of Figure~\ref{Tracks}). It is important to note that, unlike solar metallicity stars, metal-poor massive stars no longer return to the left side of the diagram, even when the effects of stellar rotation are taken into account in the models. Therefore, the most suitable point when the star may already start losing its outer shells, but is still quite hot, is the end of hydrogen combustion in the core and the beginning of the BSG stage. 
Figure~\ref{Tracks} shows the points that we took as the current evolutionary position of the central star. 
From the evolutionary tracks, we derived the chemical composition (abundances of H, He, C, N, O), log\,$g$, mass $M_*$, luminosity $L_*$, temperature $T_{\rm eff}$, and radius. We set the silicon Si, sulfur S and iron Fe abundances as 0.1 solar. 
The wind velocity was calculated using the formula from \citet{KudritzkiPuls2000}:
\begin{equation}
V_\infty=2.65\cdot V_{\rm esc}
\end{equation}
where $V_{\rm esc}$ is escape velocity. Mass loss rate $\dot{M}$ was calculated using the formula from  \citet{Vink2000}: 
%
\begin{multline}\label{mdotvink}
{\rm log}\ \dot{M} = -6.697 +2.194\ {\rm log}\ (L_*/10^5) - 1.313\ log\ (M_*/30) -\\
- 1.226\ {\rm log}\ \left( \frac{V_\infty/V_{\rm esc}}{2.0}\right)  + 0.933\ {\rm log}\ (T_{\rm eff}/ 40000) - \\
-10.92\ \{{\rm log}\ (T_{\rm eff}/ 40000)  \}^2~.
\end{multline}
%
We used these parameters (summarized in Table~\ref{tab:cmfgenparameters}) as input parameters for stellar atmospheric code {\sc cmfgen} \citep{Hillier5}. As a result, we got the model spectrum in the absolute flux units covering the whole spectral range (see Figure \ref{SEDs}). 

Stellar wind was assumed to be clumpy with an empty interclump medium \citep{Hillier99}. We calculated two models -- Model-BSG1 and Model-BSG2 -- for two different values of volume filling factor $f=1$ and $f=0.1$. Our calculation show that the value of $f$ does not affect spectral energy distribution (in exception of \HeII line intensity), and thus it is not critical for our task. 

\citet{Stromlo2021} calculated the grid of evolutionary tracks for massive stars in the wide range of  metallicities using Modules for Experiments in Stellar Astrophysics (MESA) code \citep{Mesa2011}.  \citet{Stromlo2021}  first implemented the Galactic Concordance abundances to the stellar evolution models instead of solar, scaled-solar, or alpha-element enhanced abundances. Right panel of Figure~\ref{Tracks} presents evolutionary tracks from \citet{Stromlo2021} for ${\rm Fe/H=-0.8}$. 
As it can be seen from the right panel of Figure~\ref{Tracks}, stars with the initial mass of $M_*>70\Msun$ with rotation $V/V_{\rm crit}=0.2$ after the Main Sequence move to the right and then return to the left to the WR stars region, while the tracks from \citet{GenevaTrack} do not allow creating high-mass WR stars at low  metallicity.
We calculated spectra of stars with mass $M_{\rm init}$=75 and 80~$\Msun$ with $V/V_{\rm crit}=0.2$ and 0.4. We took abundances of H, He, C, N, O, Fe, Si, S, as well as wind parameters $V_\infty$ and $\dot{M}$ from the evolutionary tracks (Table~\ref{tab:cmfgenparameters}). The filling factor was set to $f = 0.1$ for all the WR models.

One more set of evolutionary tracks has been proposed by  \citet{Szecsi2015} for low-metallicity stars with fast rotation, leading to efficient mixing and therefore chemically-homogeneous evolution. Such objects should display nearly unchanged nitrogen abundance during the stage of hydrogen burning  while having high effective temperatures \citep{Szecsi2015,Brankica2019}, thus moving towards the left part of the HR diagram. Since we see significant nitrogen overabundance in Object\,\#A relative to its host galaxy, we may conclude that the object is not undergoing this sort of chemically-homogeneous evolution. Thus, we did not consider these models in our analysis.

\subsection{{\sc cloudy}  models of the nebula}\label{mod_of_neb}

After calculation of the spectrum of the central source, we modelled the spectrum of the surrounding nebula. For that, we used the {\sc cloudy}  photoionization code \citep{Ferland1998,Ferland2017}, version 17.01 \citep{Ferland2017}, and the {\sc pyCloudy} package  \citep{2013ascl.soft04020M} that works with input and output files of {\sc cloudy} code. 

For each of the six models of central source ($\rm BSG_1$, $\rm BSG_2$, etc.) we computed a grid of  {\sc cloudy} models varying nitrogen abundance (namely, $\log({\rm N/O})$), hydrogen density, and outer radius of the nebula. We varied these parameters in the ranges as given in the  Table~\ref{tab:param}. Range for hydrogen density was chosen based on typical density of \HII regions, as well as nebulae around WR stars ($50-100$~${\text{cm}}^{-3}$).  

The stellar model parameters were used to set the luminosity, $T_{\rm eff}$, and SED of the central ionizing source. The metallicity of the nebula is set equal to one of the NGC~4068 galaxy in vicinity of the object~\#A: $Z=0.1~Z_\odot$. Table~\ref{tab:abund} shows the list of chemical abundances which were used in our calculations. The abundances, except for nitrogen and helium, are scaled by the factor 0.1 to suit the metallicity of NGC\,4068. We adopted a closed spherical geometry of the nebula with inner radius 3~pc and the volume filling factor $f=0.15$. Such value for the inner radius was chosen based on typical size of the nebulae around WR stars (see for example \citet{Stock2010}). Setting the lower inner radius ($0.3-3$~pc) doesn't affect the results, 

To estimate the parameters of the Object\,\#A, we compared {\sc cloudy} spectra with the spectrum stacked from two SCORPIO-2 spectra. This spectrum has better S/N ratio than the spectra from TDS. It also exhibits the \HeII 4686~\AA line, which is an important diagnostic line, but not detected in our TDS spectra. 

\begin{table}
\caption{Parameters of  the grid  for \Cloudy ~models of the nebula. } 
\label{tab:param}
\begin{tabular}{llll}
\toprule
                     & min value   &   max value     & step \\
\midrule               
$R_{\rm max} {\rm [pc]} $      &      6.4    &         22.6   & 3.24 - 1.62  \\
$\log(n_e)$ ${\rm  [cm^{-3}]}$  &     0.8       &        2.2   &  0.2 - 0.1 \\
${\rm log}(N/O)$          &     -1.5    &       0.6     &   0.4 - 0.1   \\
\bottomrule
\end{tabular}
\end{table}

\begin{table}
\caption{Log(X/H) set for models of the nebula, where X/H is abundance of an element relative to hydrogen. For nitrogen, the initial value is given. Galaxy metallicity value is set by Z=0.1 $\mathrm{Z_{\astrosun}}$, given values are scaled for this metallicity.}
\label{tab:abund}
\begin{tabular}{lr|lr|lr|lr}
\toprule
He  & -1.022 & C   & -4.523   & N$^*$  & -4.456 & O  & -4.398 \\
Ne  & -5.222 & Na  & -7.523   & Mg     & -6.523 & Al & -7.699 \\
Si  & -6.398 & P   & -7.796   & S      & -6.0   & Cl & -8.0   \\
Ar  & -6.523 & Ti  & -7.9788  & Fe     & -6.523 & Ni & -8.0   \\
\bottomrule
\end{tabular}
\end{table}

\begin{figure*}
		\begin{minipage}[h]{0.99\linewidth}
			\center{\includegraphics[width=17cm]{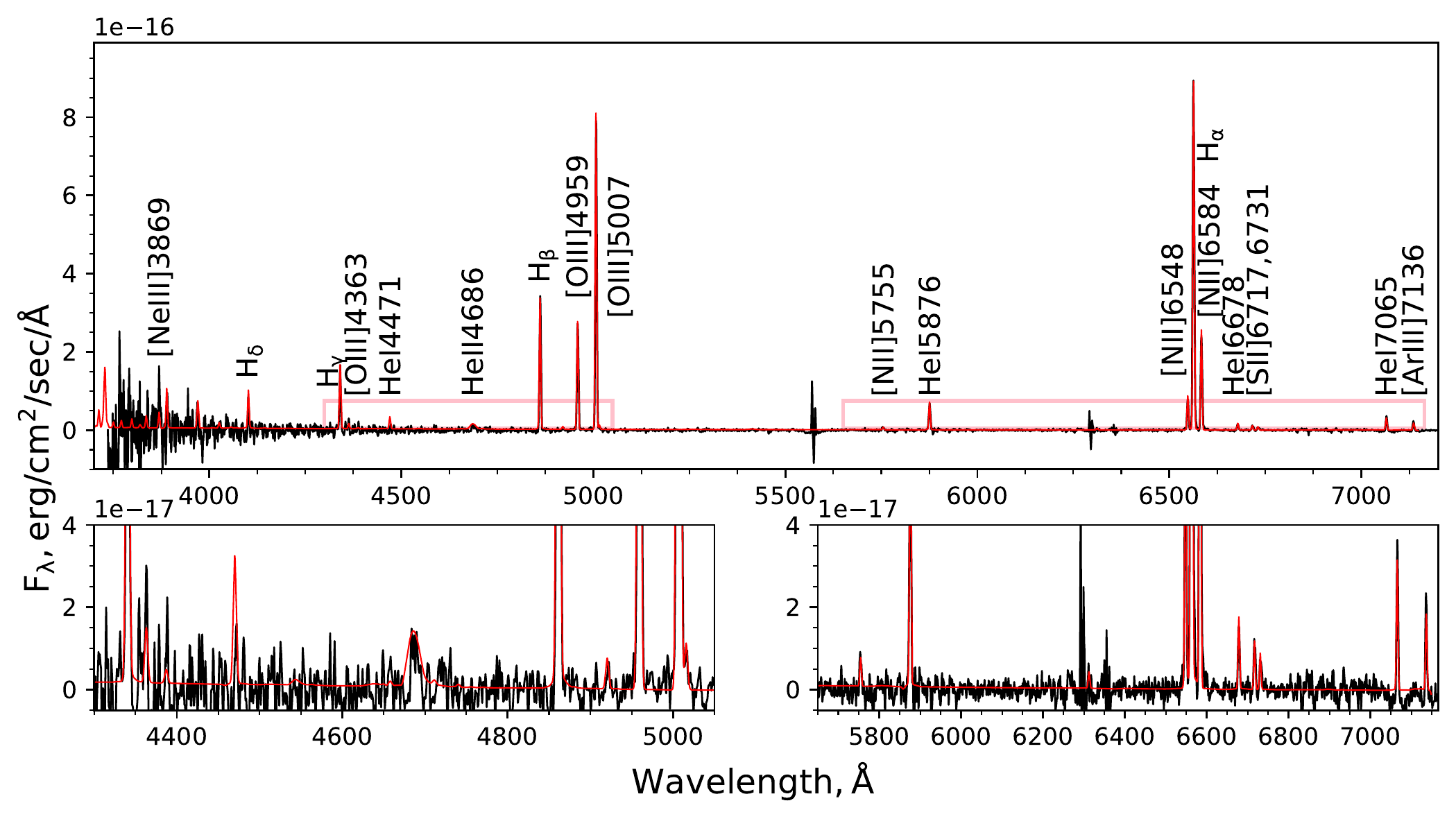}}
		\end{minipage}
		\caption{Observed SCORPIO-2 spectrum (black colour) and the best-fit {\sc cloudy}  model (red colour) convolved with a Gaussian to be visually compared with the observations. Best-fit model has the following parameters: ionizing star WR3; nebula parameters:  $\mathrm{R_{\rm max}}$ = 13 pc, $\log$(N/O)=0.15, $\mathrm{\log(n_e)}$=1.9 $\mathrm{cm^{-3}}$.
		}
		\label{comparison}
\end{figure*}

\subsection{Best-fit model of the nebula}\label{bestfit}

To choose the most successful models, we used the $\chi^2$ criterion, which takes into account the  lines listed in Table \ref{tab:chi2} using the following relation:
\small{$$\sum_{N}\chi^2=\sum_{wl \leq 5007} \left(\frac{F_o/\mathrm{H_{\beta}}_o-F_m/\mathrm{H_{\beta}}_m}{\delta(F_o/\mathrm{H_{\beta}}_o)}\right)^2+$$$$+\sum_{wl \geq 5007} \left(\frac{F_o/\mathrm{H_{\alpha}}_o-F_m/\mathrm{H_{\alpha}}_m}{\delta(F_o/\mathrm{H_{\alpha}}_o)}\right)^2, \text{where}$$}\small{$$\delta(F_o/\mathrm{H_{i}}_o)=\sqrt{\left(\delta F_o/ \mathrm{H_{i}}\right)^2+\left(\delta H_i \cdot F_o/H_i^2\right)^2}, i = \alpha, \beta,$$}$F_o$, $F_m$ are the fluxes of the emission lines in the observed and model spectra, respectively; $\mathrm{H_{\beta}}_o$, $\mathrm{H_{\alpha}}_o$ are the observed fluxes of the \Hb and \Ha lines; $\mathrm{H_{\beta}}_m$, $\mathrm{H_{\alpha}}_m$  are the modelled fluxes of these lines; N is the total number of emission lines used to compare the models. The chosen weights $\delta F_o$ and $\delta H_i$ correspond to the uncertainties of the measured fluxes (Table~\ref{tab:rel_fluxes}). The resulting value of $\chi^2$ is equal to
\small$$\chi^2 = \frac{\sum_{N}\chi^2}{N}.$$ 
Note that we exclude the \HeII $\lambda4686$~\AA\, line from the calculation of $\chi^2$ and considered it separately because it is probably associated with the stellar atmosphere, and not with the nebula (see below). 

\begin{table}
\caption{$\chi^2$ for considered lines for the best-fit {\sc cloudy} WR3-ionized  model of the nebula.}
\label{tab:chi2}
\begin{tabular}{llll}
\toprule
$\lambda$ & $\chi^2$ & $\lambda$ & $\chi^2$ \\
\midrule   
\HeII $\lambda4686$ &  0.3    & \OIII $\lambda4363$ & 0.001 \\
\OIII $\lambda5007$ &  0.02       & \HeI $\lambda5876$ &  0.5\\
\NII $\lambda6583$ & 1.1          & \HeI $\lambda6678$ & 0.4 \\
\SII $\lambda6716$ & 0.2           & \SII $\lambda6731$ &  0.004\\
\HeI $\lambda7065$ & 19.0          & \ArIII $\lambda7136$ & 10.1\\

\bottomrule
\end{tabular}
\end{table}

\begin{figure}
		\centering
		\includegraphics[width=1\linewidth]{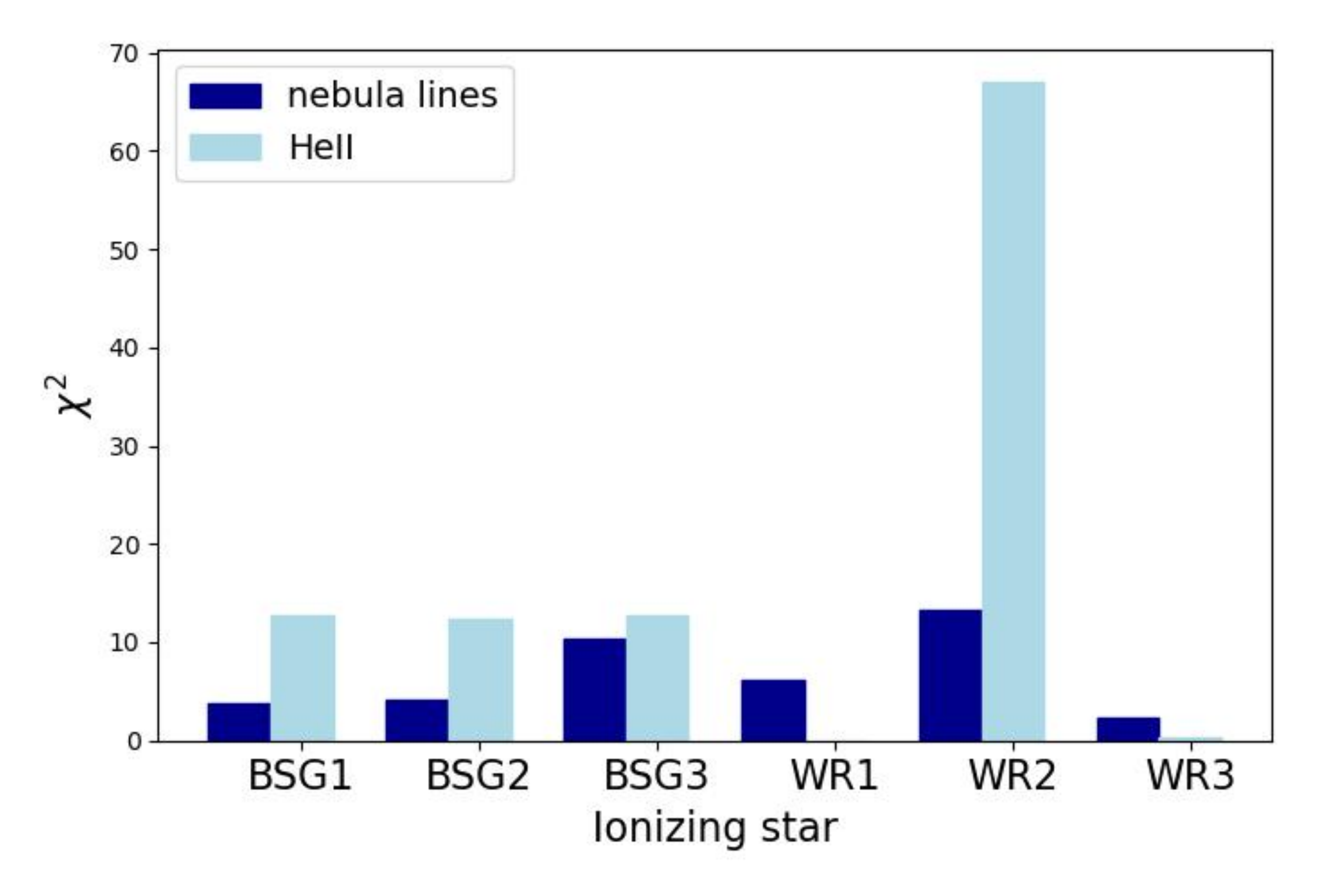}
		\caption{Comparison of the $\chi^2$ for the nebula emission lines ($\chi^2$ is normalized to the number of lines) and $\chi^2$ of \HeII line of the models of the Object \#A. BSG1, BSG2, WR1 and WR2 stars show better fitting results for the nebula emission lines than BSG3 and WR2 stars. But WR stars significantly better reproduce \HeII line, which one is emitted by the star, according to our modelling. Thus, we use WR1 and WR3 stars to estimate the parameters of the star ionizing nebula.}
		\label{Chi2}
	\end{figure}

The best agreement between the modelled spectrum and the observed one was achieved for simulated nebulae ionized by the stars BSG1, BSG2, WR3 or WR1 (see Figure~\ref{Chi2}). Among them, only the spectra of WR stars produce bright enough \HeII line. Note, however, that in all obtained \Cloudy{} models the nebular \HeII line is fainter than in the observed spectrum, while accounting for the \HeII in the incident stellar spectrum leads to a good agreement with the observations. Thus, our models demonstrate that the observed \HeII line is probably produced in the atmosphere of the ionizing star instead of the surrounding nebula. In addition, a significant nitrogen overabundance, reproduced by our models, is characteristic of WR stars, which normally eject metals into the surrounding ISM. Therefore, we chose the best model among the modelled WR stars. Note that for WR1- and WR3-ionized models of the nebula there are two model grid nodes having different log(N/O) values where $\chi^2$ converges to a minimum (see for example slices of `WR1' and `WR3' models in Figure~\ref{model_n2}). We built finer model grids for these ionizing stars considering nitrogen abundance varying around $\mathrm{log(N/O)}=0$ and obtained good agreement between our models and observed spectrum, including the flux of the \HeII line. 
Thus, we consider the models of the nebula, ionized by WR1 and WR3 stars, the best. WR1 and WR3 stars provide almost similar results. Thus, below, we will consider only a model built using WR3 star as a sample ionizing source.

$\chi^2$ values for each line of our best-fit model of the nebula ionized by WR3 star are given in the Table~\ref{tab:chi2}. The comparison of the spectrum of this model with the observed one is shown on Figure~\ref{comparison}. 
Values of the nebula model parameters with the smallest $\chi^2$ are as follows: $R_{\rm max}$ = 13 $\mathrm{pc}$, $\log(\mathrm{N/O})$=0.15, $\log(n_e)$=1.9 $\mathrm{cm^{-3}}$. This model is consistent with the nitrogen abundance derived from the observations ($\mathrm{log(N/O)} = -0.04 \pm 0.27$, see Section~\ref{est_neb}). 

The model spectrum of the nebula ionized by WR1 and WR3 star, where \HeII line is formed entirely in the atmosphere of the star, is in good agreement with the observations. For the nebula models with BSG1 or BSG2 stars as ionization sources, there is no noticeable \HeII line produced due to relatively lower helium abundances. Besides of helium fraction,  \HeII$\lambda$4686 is also sensitive to mass loss rate $\dot{M}$. Therefore, it would be possible to increase the intensity of \HeII$\lambda$4686 line in the model spectrum by increasing $\dot{M}$. In the model of BSG1 star, we counted  $\dot{M}$ according to Equation~\ref{mdotvink} from \citet{Vink2000}. However, \citet{Krticka2017} and \citet{MassLoss2020} show that mass loss rate for massive stars is usually lower than the prediction of \citet{Vink2000}. Thus, we don't have physical reasons to increase the model value of $\dot{M}$ for BSG1, and we may conclude that model WR1 and WR3 are the optimal models of ionizing source.

\section{Discussion}\label{sec:discussion}

Local dwarf irregular galaxies (dIrr) remain the best laboratories to study the formation and evolution of massive metal-poor stars due to their proximity and wide range of metallicity values. But statistics of the most massive stars (with $M > 50~ M_{\astrosun}$) does not significantly improve through the years. For now, the population of massive stars in Local Group dwarf galaxies Leo\,P, Sextans\,A, IC\,1613, Magellanic Clouds and others are studied in different works \citep[e.g.][]{2017IAUS..329..313G, Ramachandran2019, 2014A&A...572A..36T} but the most massive stars in these galaxies do not exceed the initial mass of 60~$M_{\astrosun}$ (\citet{2019arXiv190804687G}). Meanwhile, identification and subsequent analysis of the more massive stars ($\approx$100 Msun) is essential for our understanding of the evolution of metal-poor massive stars in the early Universe. 
Our results imply that the integral-field spectroscopy with high spectral resolution can be served as a tool for uncovering of such massive stars in nearby galaxies by the imprints of their feedback in the small-scale kinematics of the ionized gas. 

While both models of the central source in the object~\#A (BSG and WR) explain the observed spectrum, we are inclined to prefer the WR scenario due to the presence of \HeII line in the spectrum and the measured nitrogen overabundance. 
WR  stars form either  1) through single-star evolutionary channel (so-called Conti scenario, named after \citealt{Conti1975}) when the star loses significant part of its mass during LBV phase \citep{Conti1984}; or 2) as a stripped component in the binary system after mass transfer event \citep{Paczynski1967}. The second scenario becomes more important in low metallicity environment, where stellar winds are weaker \citep{Vink2001}. 
In our work, we used the simplest assumption that the
central source of Object\,\#A is a single evolved massive star. Good agreement between the observed and modelled spectra, one confirms evolutionary predictions by \citet{Stromlo2021} and a possibility of formation of WR stars in low metallicity environment.

Significant nitrogen overabundance,  detected in Object\,\#A, indicates that we observe the ejected stellar material rather than the swept-up surrounding gas. In this scenario, the broad component in the \Ha line profile observed just outside the object~\#A (Fig.~\ref{Ha_profile}d) can be related with the emission behind the shock front produced by the interaction of a stellar wind with the surrounding ISM. The nitrogen overabundance is often observed towards the WR stars \citep[e.g.][and references therein]{PM2013}, and the resolved studies of the Galactic WR nebulae often associate a high N/O with the stellar ejecta \cite[e.g.][]{Stock2011, Esteban2016, Fernandez-Martin2012}. These studies often claim the slight deficient of the Ne/O abundance, which is not seen in our spectra of the object~\#A. 

Interestingly to note that the spectrum of the low-ionization area of the X-ray emitting zone at the periphery of the WR nebula NGC\,6888 around the star WR\,136 exhibits the same peculiar ratio of [S~\textsc{ii}]/[N~\textsc{ii}] line fluxes as in our Object~\#A (see fig.~10 in \citealt{Fernandez-Martin2012}), though they don't report the detection of \HeII $\lambda$4686~\AA\, line, and the \OIII lines are significantly fainter than in our case. In that sense, it more resembles the mentioned earlier PN in Sextans~B \citep{2005AJ....130.1558K}. The spectrum of the studied object~
\#A exhibits several unusual features making it a unique object, and we still haven't found any sibling to this in the published literature.

We mention here also possible spectral variability of the object~\#A. In this work, we used several spectra from different instruments obtained with intervals of several years. As it can be seen from Table~\ref{tab:rel_fluxes}, the reported fluxes ratios of the low- and high-excitation lines to the Balmer lines (namely, \NIIHa and \OIIIHb) are not conserved, and the difference is above the uncertainties. 
In order to assess whether this difference may be due to instrumental effects, or due to the non-uniform contribution of the diffuse ionized gas, we performed in December 2020 several additional observations with TDS at the 2.5-m telescope in CMO with different position angles of the slit (see Section~\ref{longslit_spectroscopic_obs}). All individual exposures reveal the same flux ratios as in the TDS spectrum in Fig.~\ref{spectrum_CMO} and in Tab.~\ref{tab:rel_fluxes}, independently of the orientation of the slit. Two SCORPIO-2 spectra obtained with interval of 1 year are also identical. Taking into account that the SCORPIO-2 data from the BTA 6-m telescope were obtained $5-6$ years earlier, 
one may suspect that the Object\,\#A is variable on such timescale, but further observation are necessary to confirm this. Also  NGC~4068 was recently observed with SparsePak IFU \citep{Hunter2022}, and these data could help to verify a probable variability of Object\,\#A, but unfortunately, this object fell into the gap between the fibers and thus was not detected in these data.
Spectral variability is typical for evolved massive stars. Variability of hydrogen lines was registered for many BSGs surrounded by circumstellar nebulae (see for example \citealt{Hendry2008, Gvaramadze2015, Gvaramadze2018}).  Although WR stars are not usually showing variability \citep{Crowther2007}, some LBVs in minimum of brightness during hot phase show WR spectra (for example AG\,Car \citep{groh}, Romano's star \citep{MaryevaGalaxies}). Therefore, we cannot exclude that the Object\,\#A belongs to the class of LBVs. If Object\,\#A is LBV then it is one of the brightest LBVs in nearby galaxies along with NGC\,2363-V1 \citep{Drissen2001, Petit2006}.

\section{Summary}\label{sec:summary}
In this paper, we present the analysis of the unusual stellar-like emission-line object at the outskirt of the nearby ($D\sim4.36$~Mpc) low-metallicity ($Z\sim0.1Z_\odot$) dwarf galaxy NGC~4068. This object~\#A
was uncovered by its high velocity dispersion of the ionized gas in \Ha line, and subsequent spectral observations revealed its peculiar emission spectrum -- very faint \SII~$\lambda6717,6731$\AA\, lines, bright \NII$\lambda6548, 6584$\AA\ and \OIII$\lambda4959, 5007$\AA\, lines, and presence of \HeII$\lambda4868$\AA\, emission. We investigate this object based on our new long-slit spectral observations performed at the 6-m BTA (SAO RAS) and the 2.5-m CMO (SAI MSU) telescopes. We also rely on our data obtained earlier with Fabry-Perot interferometer at the 6-m BTA telescope and on the archival \HST imaging. From the analysis of the observational data, we derived the following properties of the object~\#A:

\begin{itemize}
    \item Central ionizing source is a single non-resolved in \HST data object with effective temperature $T_eff \sim 40$~kK and bolometric luminosity $L_* \sim (1.5-1.7)\times10^6 L_\odot$
    \item Central source is surrounded by the nebula unresolved in our images and spectra. We estimated the upper limit of its radius $R < 15$~pc and the electron density $n_e \sim 120$~cm$^{-3}$. The nebula is expanding with velocity $V_\mathrm{exp} \sim 38\ \kms$, and the upper limit of its kinematical age is $t\sim 0.5$~Myr.
    \item The oxygen abundance of object~\#A $12+\log({\rm O/H}) \sim 7.4-7.5$ in agreement with metallicity in that area of NGC~4068, and relative abundances of Ar/O and Ne/O are consistent with the metallicity. Meanwhile, we found strong overabundance by nitrogen: $\log({\rm N/O})= -0.04\pm0.27$, that is more typical for a solar metallicity.  
    \item Observed properties of the object~\#A imply that it is a single massive metal-poor star at the stage producing powerful stellar wind. The nebula probably represents expanding stellar ejecta driven by the stellar wind. The signs of its interaction with the ISM are detected in a form of a broad underlying component in \Ha line profile close to the object~\#A. 
    \item Changes of the \OIIIHb\, and \NIIHa\, line ratios on the scale of 5 years points to the probable variability of the object~\#A
\end{itemize}

We calculated a set of the \textsc{Cloudy} + \textsc{cmfgen} models aimed to reproduce the observed peculiarities in the spectrum of object~\#A. We considered several scenarios of the central ionizing source corresponding to an evolved massive star of ($M_*=75-80~\mathrm{M_{\astrosun}}$), which already reached BSG or WR stage and has a strong stellar wind. For modelling the WR star at such a low metallicity, we assumed the \citet{Stromlo2021} evolutionary tracks. The model corresponding to the nebula ionized by WR star yielded a best agreement with the observed spectra -- fluxes in all main emission lines were successfully reproduced. The best-fit model of the ionizing star and nebula yielded parameters close to that derived in the observations: $L_* \sim 2\times10^6\Lsun$, $R \sim 13$~pc, $n_e \sim 80$~cm$^{-3}$, $\log({\rm N/O}) \sim 0.15$.
As follows from our models,  the emission in \HeII$\lambda4686$~\AA\, is probably produced in the atmosphere of the star, not in the nebula. Given the probable variability of the spectrum, we cannot exclude that the object~\#A is LBV star in a hot phase.

Our work demonstrates the potential of the high spectral resolution integral field spectroscopy in the identification of the massive stars by the imprints of their feedback in the kinematics of the ionized gas. The conclusions about the central star were obtained based mostly on the optical spectrum of a nebula, and thus further confirmation of our findings with UV spectroscopy or X-ray observations are desired. If confirmed, the object~\#A appears to be one of the most massive and luminous low-metallicity massive star (WR or LBV) found up to date in the nearby galaxies. Massive stars in low metallicity galaxies are unique targets for testing our models of stellar evolution. In particular, our results support the predictions \citet{Stromlo2021} about the possibility of the formation of luminous WR stars at low metallicity.

\section*{Acknowledgements}\label{sec:acknow}

The authors thank S. Zheltoukhov for his assistance in the CMO 2.5-m observations. We also thank D. Bomans., A.Yu.~Knizev, and T.A. Lozinskaya for useful discussion and advises.

O.M.  acknowledges the support from European Union's Framework Programme for Research and Innovation Horizon 2020 (2014-2020) under the Marie Sk\l{}odowska-Curie Grant Agreement No. 823734, and the project RVO:67985815 of the Academy of Sciences of the Czech Republic. OE acknowledge funding from the Deutsche Forschungsgemeinschaft (DFG, German Research Foundation) in the form of an Emmy Noether Research Group (grant number KR4598/2-1, PI Kreckel). 
This study   based on the data obtained at the   unique scientific facility   the Big Telescope Alt-azimuthal SAO RAS and  was supported  under  the   Ministry of Science and Higher Education of the Russian Federation grant  075-15-2022-262 (13.MNPMU.21.0003). 
This research made use of Astropy (\url{http://www.astropy.org}) a community-developed core Python package for Astronomy \citep{astropy:2013, astropy:2018}. 
Based on observations made with the NASA/ESA Hubble Space Telescope, and obtained from the Hubble Legacy Archive, which is a collaboration between the Space Telescope Science Institute (STScI/NASA), the Space Telescope European Coordinating Facility (ST-ECF/ESA) and the Canadian Astronomy Data Centre (CADC/NRC/CSA). 

\section*{DATA AVAILABILITY}
The data underlying this article will be shared on reasonable request to the corresponding author.

\bibliographystyle{mnras}
\bibliography{N4068_theStar} 

\appendix
\section{Appendix A: Model grids}

\begin{figure*}
		\begin{minipage}[h]{0.99\linewidth}
			\center{\includegraphics[width=17cm]{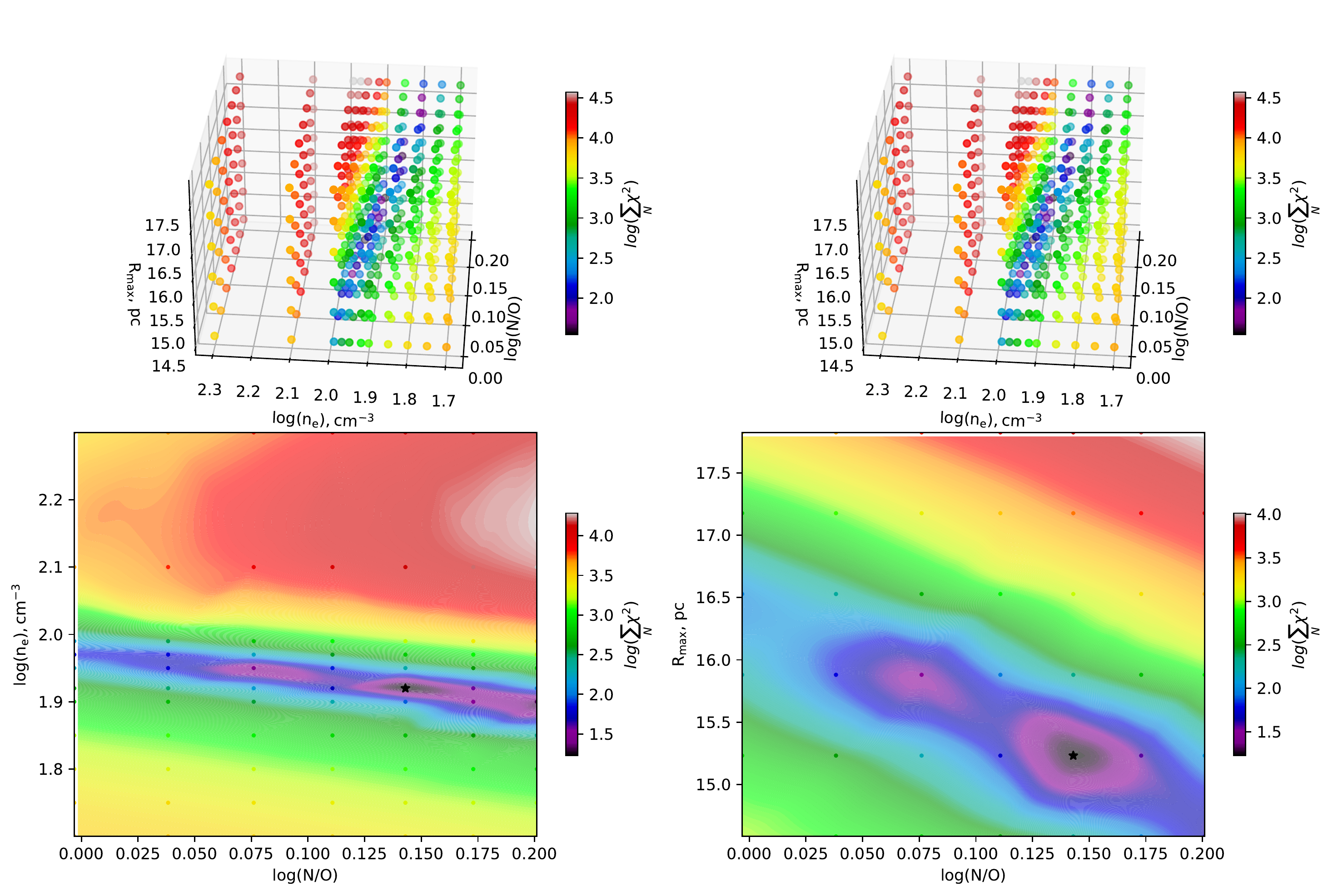}}
		\end{minipage}
		\caption{Grids of \Cloudy\, models for WR3-ionized nebula with different resolution. Top panel: right-hand plot shows the grid with low precision, left-hand plot shows the finer grid in the parameter space around the minimum value of $\chi^2$. Bottom panel: the plots show slices of the finer model grid that cross the minimum value of $\chi^2$. Color bar corresponds to the $\log(\sum_{N}\chi^2$).
		}
		\label{WR3_net}
\end{figure*}

\begin{figure*}
		\begin{minipage}[h]{0.96\linewidth}
			\center{\includegraphics[width=16cm]{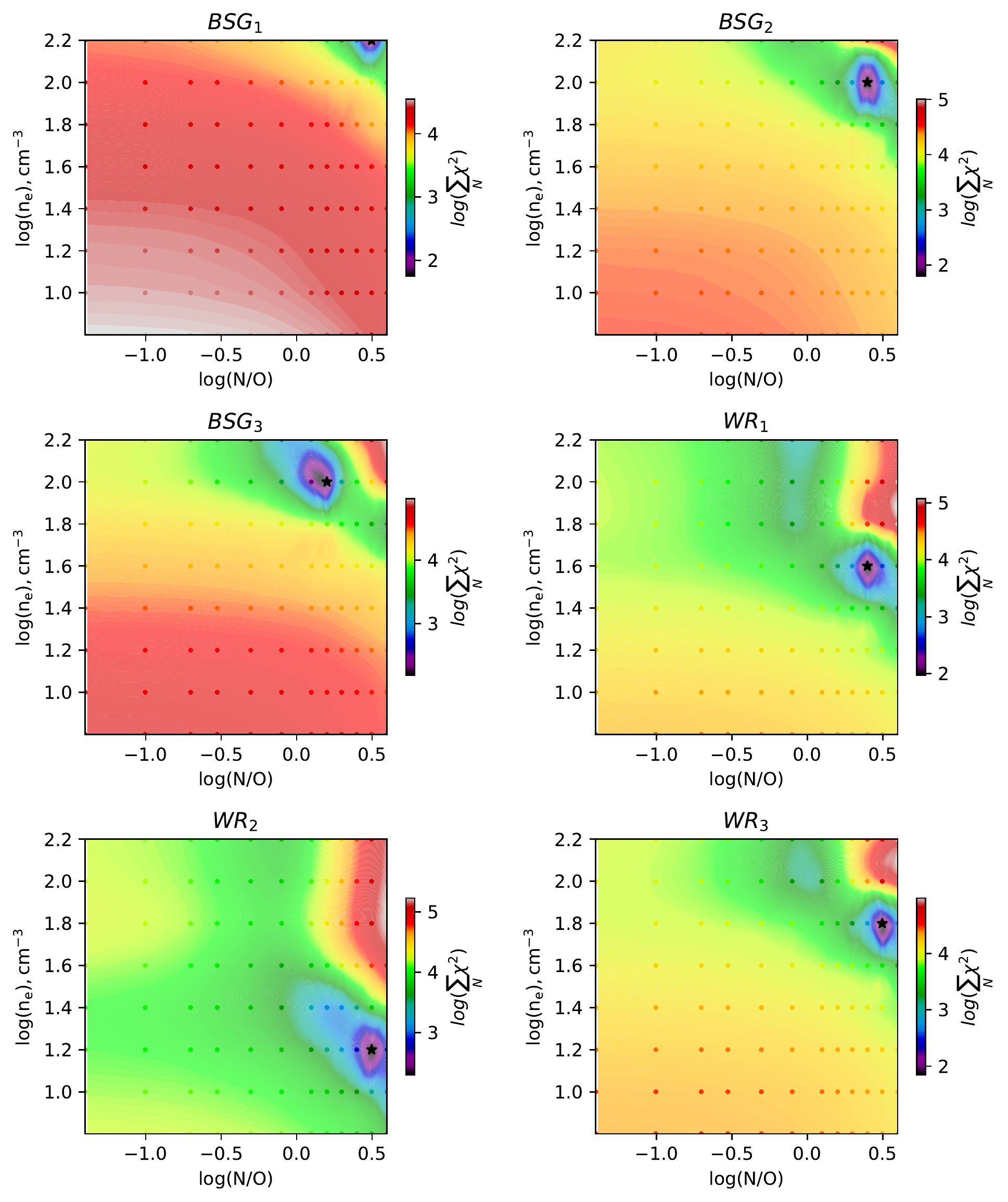}}
		\end{minipage}
		\caption{Slices of the low-resolution \Cloudy\, model grids. Color bar corresponds to the $\log(\sum_{N}\chi^2$).
		}
		\label{model_n1}
\end{figure*}

\begin{figure*}

		\begin{minipage}[h]{0.96\linewidth}
			\center{\includegraphics[width=16cm]{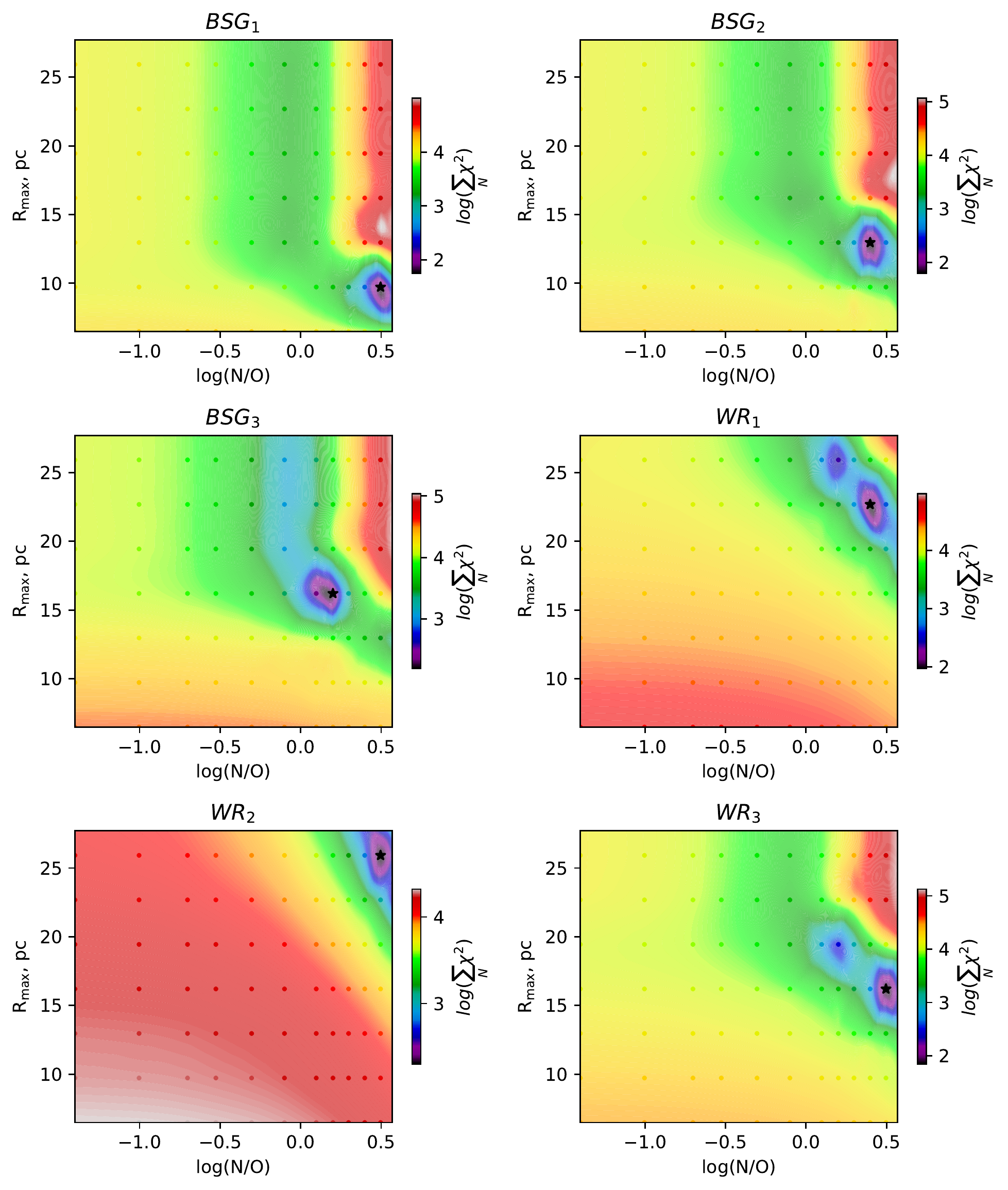}}
		\end{minipage}

		\caption{Slices of the low-resolution \Cloudy\, model grids. Colour bar corresponds to the $\log(\sum_{N}\chi^2)$. All of our models have better agreement with observations with the highest values of $\log({\rm N/O})$. Note that for WR1- and WR3-ionized models of the nebula there are two model grid nodes having different $\log({\rm N/O})$ values where $log(\sum_{N}\chi^2)$ converges to a minimum. 
		}
		\label{model_n2}
\end{figure*}

Slices of model grids of \Cloudy~ models of the Object \#A are presented. Figure \ref{WR3_net} shows the grids of the best-fit model of the nebula, ionized by WR3 star. Figures \ref{model_n1} and \ref{model_n2} show all the models built in wide parameters range.

\bsp	
\label{lastpage}
\end{document}